\documentclass[nofootinbib,a4paper,12pt]{article}
\usepackage[english]{babel}
\usepackage{jheppub}
\usepackage[T1]{fontenc}
\usepackage[utf8]{inputenc}
\usepackage{float}
\usepackage{slashed}
\usepackage{multicol,multirow}
\usepackage{amssymb}
\usepackage{amsmath}
\usepackage{subcaption}
\usepackage{array}
\usepackage[dvipsnames]{xcolor}
\usepackage{hyperref}
\hypersetup{colorlinks=true}
\usepackage{longtable}
\usepackage{graphics,graphicx,epsfig,ulem,booktabs}
\numberwithin{equation}{section}
\usepackage{url}
\usepackage{physics}
\usepackage{graphicx}
\usepackage{simpler-wick}
\usepackage{feynmp}
\usepackage[compat=1.1.0]{tikz-feynman} 
\usepackage{bm}
\usepackage[fleqn]{nccmath}
\usepackage{bbold}
\usepackage{kantlipsum}
\usepackage{braket}
\usepackage{empheq}
\usepackage[most]{tcolorbox}
\usepackage{caption}
\usepackage{subcaption}
\usepackage{fancyhdr} 
\usepackage{longtable}
\usepackage{pgfplots}
\usepackage[nameinlink,capitalize]{cleveref}
\usepackage{orcidlink}

\newcolumntype{C}[1]{>{\centering\let\newline\\\arraybackslash\hspace{0pt}}m{#1}}

\makeatletter
\AddToHook{cmd/appendix/before}{\def\cref@section@alias{appendix}}
\makeatother
\pgfplotsset{compat=1.18}

\hypersetup{%
  colorlinks=true, linktocpage=true, pdfstartpage=1, pdfstartview=FitV,
  breaklinks=true, pageanchor=true,
  pdfpagemode=UseNone,
  plainpages=false, bookmarksnumbered, bookmarksopen=true, bookmarksopenlevel=1,
  hypertexnames=true, pdfhighlight=/O,
  urlcolor=RoyalBlue, linkcolor=ForestGreen, citecolor=RoyalBlue,
  pdftitle={B-meson decay width up to 1/mb3 corrections within and beyond the Standard Model},
  pdfauthor={Martin Lang, Alexander Lenz, Ali Mohamed, Maria Laura Piscopo, Aleksey V. Rusov},
  pdfsubject={},
  pdfkeywords={},
  pdfcreator={pdfLaTeX},
  pdfproducer={LaTeX with hyperref}
}

\title{\boldmath 
$B$-meson decay width up to $1/m_b^3$ corrections within and beyond the Standard Model
}

\preprint{\small TUM-HEP-1583/25, P3H-25-110, SI-HEP-2025-29, Nikhef~2025-020}

\author[a]{Martin Lang\orcidlink{0000-0002-8610-1376}}
\author[a]{, Alexander Lenz\orcidlink{0000-0003-3976-035X}}
\author[a]{, Ali Mohamed\orcidlink{0009-0006-3466-2709}}
\author[b,c]{, Maria Laura Piscopo\orcidlink{0000-0003-0796-5561}}
\author[d]{, Aleksey~V.~Rusov\orcidlink{0000-0002-7037-6391}}

\affiliation[a]{Physik Department, Universit\"{a}t Siegen, Walter-Flex-Str. 3, D-57068 Siegen, Germany}
\affiliation[b]{Nikhef, Science Park 105, NL-1098 XG Amsterdam, Netherlands}
\affiliation[c]{Department of Physics and Astronomy, Vrije Universiteit Amsterdam, NL-1081 HV Amsterdam, Netherlands}
\affiliation[d]{Physik Department T31, 
Technische Universit\"at M\"unchen, James-Franck-Straße 1, D-85748 Garching, Germany}

\emailAdd{martin.lang@uni-siegen.de}
\emailAdd{alexander.lenz@uni-siegen.de}
\emailAdd{ali.mohamed@uni-siegen.de}
\emailAdd{mpiscopo@nikhef.nl}
\emailAdd{aleksey.rusov@tum.de}

\abstract{
Starting from the most general effective $\abs{\Delta B} = 1$ Hamiltonian describing non-leptonic $b$-quark decays $b\to q_1 \bar q_2 q_3$, we compute analytic expressions for all matching coefficients of the two-quark operator contributions in the heavy quark expansion~(HQE) of a $B$ meson, up to mass-dimension-six. In addition, we calculate the weak-annihilation contributions, which enter the matching of four-quark operators in the HQE at dimension-six and were previously missing.
Our results
complete the calculation of beyond Standard Model (BSM) effects in non-leptonic, tree-level, $b$-quark decays relevant for $B$ meson lifetimes and lifetime ratios such as $\tau(B^0_s)/\tau(B^0_d)$. Such BSM contributions naturally arise in generic extensions of the Standard Model (SM) that aim to address the observed tensions between experimental measurements and theoretical predictions based on QCD factorisation in several colour-allowed non-leptonic $B$-meson decays. 
As a by-product of our calculation, we also determine the matching coefficients in the HQE induced by the QCD-penguin operators within the SM, including both the interference between current-current and penguin operators and the contributions quadratic in the penguin operators.    
Owing to the suppression of the QCD-penguin Wilson coefficients within the SM, these effects are typically regarded as corrections of order $\alpha_s$ and $\alpha_s^2$ in the strong coupling, respectively.
Our results reproduce the known expressions at dimension-three and provide new results for the coefficients of the chromomagnetic operator at dimension-five and of the Darwin operator at dimension-six. 
}

\begin{document}

\maketitle
\flushbottom


 \thispagestyle{empty} 

\pagebreak

\section{Introduction}
\label{sec:Intro}
The lifetime is among the most fundamental properties of an elementary particle. Lifetimes of heavy hadrons containing a $b$ quark have by now been measured very precisely at experiments like 
ALEPH, ATLAS, BABAR, BELLE, CDF, CMS, D0, DELPHI, L3, LHCb, OPAL and SLD, see e.g.\ the HFLAV review \cite{HeavyFlavorAveragingGroupHFLAV:2024ctg}. Currently, for the $B^0_d, B^+,$ and $B_s^0$ mesons, the experimental uncertainties have reached the per-mille level, namely
\begin{eqnarray}
\tau (B_d^0)^{\rm Exp.} = 1.517(4) \, \mbox{ps} \, ,
\hspace{0.5cm}  
\tau (B^+)^{\rm Exp.} = 1.638(4) \, \mbox{ps} \, ,
\hspace{0.5cm}  
\tau (B_s^0)^{\rm Exp.} = 1.516(6) \, \mbox{ps} \,,
\end{eqnarray}
whereas uncertainties for $b$-baryon lifetimes are generally larger, although these are also by now experimentally very
well established:
\begin{align}
\tau (\Lambda_b)^{\rm Exp.} & = 1.468(9) \, \mbox{ps} \,, 
& \tau (\Xi_b^-)^{\rm Exp.} & = 1.578(21) \, \mbox{ps} \,, \nonumber
\\[2mm]
\tau (\Xi_b^0)^{\rm Exp.} & = 1.477(32) \, \mbox{ps} \, ,
& \tau (\Omega_b^-)^{\rm Exp.} & = 1.64(16) \, \mbox{ps} \,.
\end{align}
The increasing experimental precision must clearly be matched by corresponding advances on the theoretical side. Within the Standard Model (SM), lifetimes of heavy hadrons are computed using the 
framework of the heavy quark expansion (HQE)~\cite{Shifman:1986mx}.\footnote{See Ref.~\cite{Albrecht:2024oyn} for a recent review and Ref.~\cite{Lenz:2014jha} for a review of the historical development.}
The current state of the art of HQE predictions for $B$ meson lifetimes is summarised in Ref.~\cite{Egner:2024lay}:
\begin{eqnarray}
\tau (B_d^0)^{\rm HQE} = 1.57^{+0.10}_{-0.06}  \, \mbox{ps} \, ,
\hspace{0.5cm}  
\tau (B^+)^{\rm HQE} = 1.70^{+0.11}_{-0.07}  \, \mbox{ps} \, ,
\hspace{0.5cm}  
\tau (B_s^0)^{\rm HQE} = 1.59^{+0.10}_{-0.06}  \, \mbox{ps} \, .
\end{eqnarray}
These values are in very good agreement with the corresponding measurements, although the theoretical uncertainties are still about a factor of 20 larger.
As it will be explained below, ratios of heavy-hadron lifetimes can be predicted with significantly higher precision. In particular, for the $B$ mesons we have: 
\begin{align}
\frac{\tau (B^+)}{\tau (B_d^0)}^{\rm Exp.} &=
1.076(4) \, , 
&
\frac{\tau (B_s^0)}{\tau (B_d^0)}^{\rm Exp.} &=
1.0021(34) \, , 
\\
\frac{\tau (B^+)}{\tau (B_d^0)}^{\rm HQE} &=
1.081^{+0.014}_{-0.016} \, , 
&
\frac{\tau (B_s^0)}{\tau (B_d^0)}^{\rm HQE} &=
1.013(7) \, .
\end{align}
We find an excellent agreement and, for these ratios, the experimental precision is only a factor 2--4 higher than the theoretical one.
In the case of baryons, not all available SM contributions have been included so far, resulting in less precise predictions than in the meson sector. The most up-to-date phenomenological analysis~\cite{Gratrex:2023pfn} yields,
\begin{align}
\tau (\Lambda_b)^{\rm HQE} & = 
1.49^{+0.18}_{-0.21}  \, \mbox{ps} \, ,
& \tau (\Xi_b^-)^{\rm HQE} & = 1.61^{+0.19}_{-0.23} \, \mbox{ps} \, ,
\nonumber \\[2mm]
\tau (\Xi_b^0)^{\rm HQE} & = 1.49^{+0.18}_{-0.21} \, \mbox{ps} \,,
& \tau (\Omega_b^-)^{\rm HQE} & = 1.69^{+0.23}_{-0.26} \,\mbox{ps}\,,
\end{align}
which agrees well with the measurements albeit with larger uncertainties.\\
Besides providing a test of our ability to predict experimental observables within the SM, heavy-hadron lifetimes can also be used to constrain the possible size of general beyond the Standard Model~(BSM) effects in non-leptonic, tree-level, heavy-quark decays. In the case of the $b$ quark, such effects are partly motivated by the observation that experimental  branching ratios for colour-allowed non-leptonic $B$-meson decays, such as 
$\bar{B}^0_s \to D_s^+ \pi^-$, deviate significantly from the theoretical predictions \cite{Bordone:2020gao}. The latter were obtained within the framework of QCD factorisation \cite{Beneke:1999br, Beneke:2000ry, Beneke:2001ev} using next-to-next-to leading order (NNLO) QCD results at leading power~\cite{Huber:2016xod} and a first estimate of next-to-leading power effects. \\
This finding has triggered significant interest in the literature: 
on the one hand, missing contributions within the SM that could explain the discrepancies have been investigated, including QED corrections~\cite{Beneke:2021jhp}
and rescattering contributions~\cite{Endo:2021ifc}. However, the latter do not appear sufficient to generate sizeable effects. 
An interesting approach is the use of light-cone sum rules~\cite{Balitsky:1989ry} for the description of the decay amplitude of non-leptonic, two-body, heavy-meson decays.
First results for the tree-level decays subject to tension, namely $\bar{B}^0_s \to D_s^+ \pi^-$ and $\bar{B}_d^0 \to D^+ K^-$~\cite{Piscopo:2023opf}, but also for two-body $D^0$ decays into $K^+K^-$ or $\pi^+ \pi^-$ final states \cite{Lenz:2023rlq,Piscopo:2024wpd}, are promising, but still suffer from very large uncertainties. Hence, further theoretical efforts are required.
Finally, final-state interactions have been studied in Ref.~\cite{Panuluh:2024cxc}.\\
On the other hand, the possibility of BSM effects in the non-leptonic tree-level decays $b \to c \bar{u} d $ and $b \to c \bar{u} s $ was investigated in e.g.\ Refs.~\cite{Meiser:2024zea,Fleischer:2021cct, Fleischer:2021cwb}.
Such BSM contributions could originate, for instance, from new left-handed $W'$ bosons~\cite{Iguro:2020ndk},
di-quark contributions~\cite{Crivellin:2023saq}, or from two-Higgs-doublet models~(2HDM), see e.g.\ Ref.~\cite{Atkinson:2021eox}.
Potential constraints on sizable BSM contributions to tree-level $b$-quark decays from collider bounds were discussed in Ref.~\cite{Bordone:2021cca} and further analysed in Ref.~\cite{Atkinson:2024hqp}, while a complementary experimental test of such BSM effects, in the case that they are CP violating, using flavour-specific CP asymmetries, was studied in Ref.~\cite{Gershon:2021pnc}. 
Phenomenological constraints derived from lifetimes and lifetime ratios have also been studied for generic BSM effects in tree-level $b$-quark decays in Refs.~\cite{Lenz:2022pgw,Bobeth:2014rda,Brod:2014bfa,Lenz:2019lvd}. In the specific case of $b \to c \bar{c} s$ transitions, these effects were explored in connection with an interesting interplay between $B_s^0$ mixing and the $b \to s \ell \ell$ anomalies~\cite{Jager:2017gal,Jager:2019bgk}. Constraints from lifetimes have furthermore been considered for semileptonic decays such as $b \to s \tau \tau$~\cite{Bordone:2023ybl}. 
We note here that most of these previous analyses of lifetimes (or equivalently total decay rates) 
relied on simplifying assumptions. In particular, studies employing the lifetime ratio $\tau(B_s^0)/\tau(B_d^0)$ did not include potentially sizeable SU(3)$_F$-breaking effects arising from dimension-six contributions.\\
In this work, we study the effect of these potential BSM contributions using a model-independent approach.
Starting from the most general effective Hamiltonian describing the non-leptonic $b$-quark decays $b \to q_1 \bar q_2 q_3$~\cite{Jager:2017gal,Jager:2019bgk,Cai:2021mlt}, with $q_1, q_2 = u, c,$ and $q_3 = d,s,$ we compute analytic expressions at LO in QCD for all matching coefficients of two-quark operators in the HQE of a $B$ meson, up to mass-dimension-six, including the leading-power results at dimension-three and the coefficients of the chromomagnetic and Darwin operators at dimension-five and six, respectively. 
At dimension-six, the Darwin operator mixes with four-quark operators under renormalisation, and a proper treatment of this mixing is required to subtract the infrared (IR) divergences that appear in the Darwin coefficients from the emission of a soft gluon from one of the light-quark propagators $q = u,d,s$, as discussed e.g.\ in Refs.~\cite{Lenz:2020oce, Mannel:2020fts}. To regularise these divergences, we consider both a finite light-quark mass and dimensional regularisation, providing an independent and crucial cross-check of the calculation.\\
The subtraction of all IR-divergent terms additionally requires the computation of weak-annihilation contributions, which enter the matching of four-quark operators at dimension-six and were previously missing. Our results thus
complete the calculation of BSM effects in non-leptonic, tree-level, $b$-quark decays relevant for $B$ meson lifetimes and lifetime ratios such as $\tau(B^0_s)/\tau(B^0_d)$. In fact, corresponding results at dimension-six for the matching of four-quark operators due to the weak-exchange and Pauli interference contributions were already computed in Refs.~\cite{Lenz:2022pgw,Jager:2017gal,Jager:2019bgk}, while recently, the leading-power results for the $b \to c \bar u d$ transition were derived in Ref.~\cite{Meiser:2024zea}. Moreover, hadronic matrix elements of dimension-six four-quark operators with generic BSM Dirac structures have been determined with a three-loop sum rule, in the heavy-quark effective theory (HQET) limit, in Ref.~\cite{Black:2024bus}. \\
As a by-product of our calculation, we also determine the SM matching coefficients up to dimension-six originating from the QCD-penguin operators, including both the interference between current-current and penguin operators and the contributions quadratic in the penguin operators.    
Owing to the suppression of the QCD-penguin Wilson coefficients within the SM, these effects are typically regarded as corrections of order $\alpha_s$ and $\alpha_s^2$ in the strong coupling, respectively.
Our results reproduce the known expressions at dimension-three and provide new results for the coefficients of the chromomagnetic operator at dimension-five and of the Darwin operator at dimension-six. \\
The paper is organised as follows: in 
\cref{sec:frame} we describe the theoretical framework, starting with a discussion of the HQE in \cref{subsec:HQE}, followed by the generalised effective Hamiltonian in \cref{subsec:Heff} and concluding with the details of our calculation, performed using the light-quark mass as an IR regulator, in \cref{Calsteps}.  Our main results are presented in \cref{Sec:Results}, with the coefficients up to dimension-six for the $b \to c \bar u d$ mode given in \cref{sec:bcud} and those for $b \to c \bar c s$ in \cref{sec:bccs}. We conclude in \cref{Sec:outlook}. 
In the appendix we list additional results. Specifically, in \cref{sec:DR} we provide the details of the independent calculation performed using dimensional regularisation; the results for the matching coefficients for the $b \to u \bar c s$ and $b \to u \bar u d$ modes are given in \cref{sec:ucs,sec:uud}, respectively; in \cref{sec:WA} we present the new expressions for the weak-annihilation contributions at dimension-six; finally, our results for the contributions of the QCD-penguin operators within the SM are discussed in \cref{sec:SM_penguin}.
\section{Theoretical Framework} 
\label{sec:frame}
\subsection{Heavy Quark Expansion }
\label{subsec:HQE}
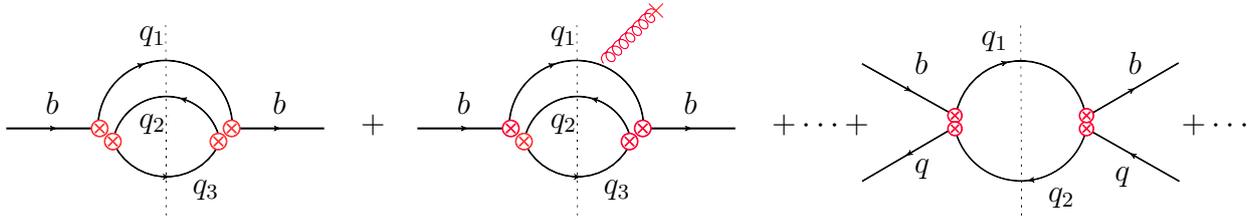
\begin{figure}[t!]
  \centering
  \setlength{\unitlength}{1pt}
  \begin{tabular}{ccccc}
    \begin{picture}(150,120)
      \put(0,0){\includegraphics[width=0.25\textwidth]{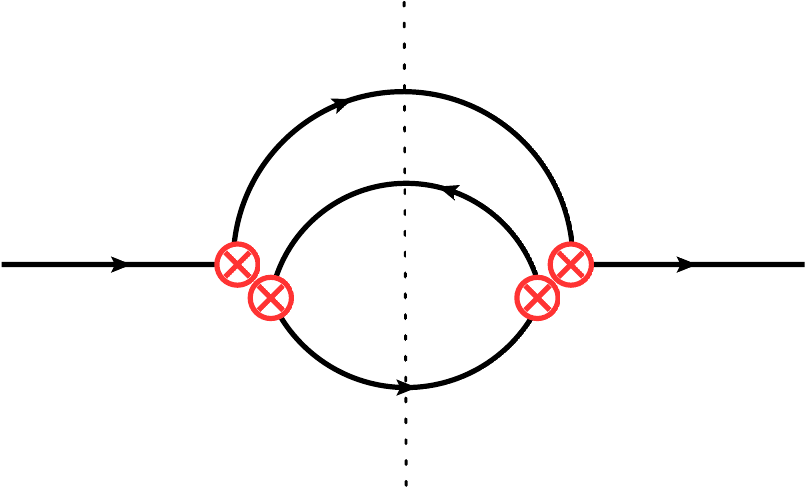}}
      \put(50,68){$q_1$}
      \put(50,35){$q_2$}
      \put(70,10){$q_3$}
      \put(15,40){$b$}
      \put(100,40){$b$}
       \put(133,33){$+$}
    \end{picture}
    \begin{picture}(150,120)
      \put(0,0){\includegraphics[width=0.25\textwidth]{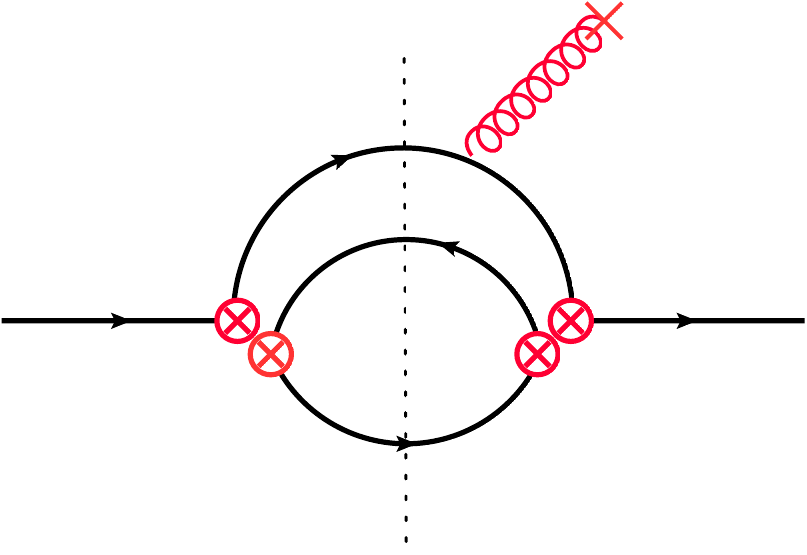}}
      \put(50,68){$q_1$}
      \put(50,35){$q_2$}
      \put(70,10){$q_3$}
      \put(15,40){$b$}
      \put(100,40){$b$}
      \put(133,33){$+ \cdots +$}
    \end{picture}
\hspace{0.3cm}
    \begin{picture}(150,120)
      \put(0,0){\includegraphics[width=0.25\textwidth]{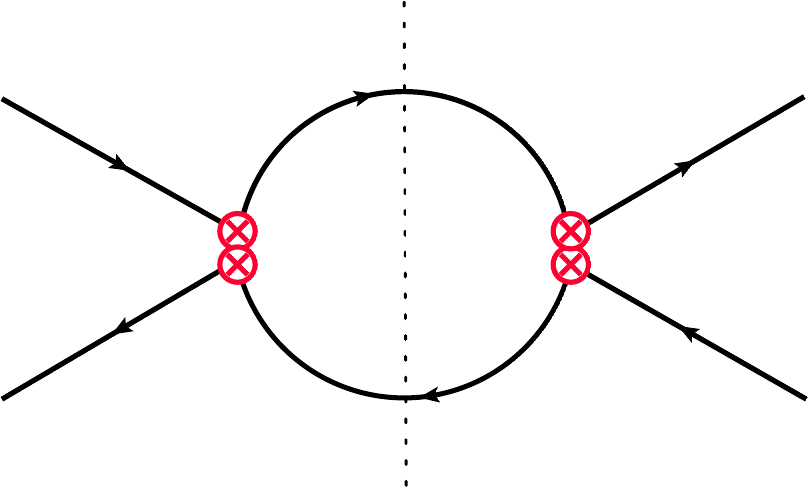}}
      \put(45,66){$q_1$}
      \put(70,7){$q_2$}
      \put(20,55){$b$}
      \put(100,55){$b$}
      \put(20,18){$q$}
      \put(95,15){$q$}
         \put(120,33){$+ \cdots $}
    \end{picture}
  \end{tabular}
\caption{Schematic representation of the HQE for $B$-meson decays due to non-leptonic $b\to q_1 \bar q_2 q_3$ decays, including two-quark contributions (two-loop) and four-quark contributions (one-loop). The crossed red vertices indicate the insertion of the $\abs{\Delta B}=1$ effective Hamiltonian. The imaginary part is taken across the dotted line.}
\label{fig:HQEDiags}
\end{figure}

In the SM, the non-leptonic decays of a $b$ quark are described by the effective Hamiltonian~\cite{Buchalla:1995vs}:
\begin{align}
{\cal H}_{\rm eff}^{\rm SM} = 
\frac{4 G_F}{\sqrt{2}} \sum_{q_3 = d, s} 
\Bigg\{ \sum_{q_1,q_2 = u,c } &
V_{q_1 b}\, V_{q_2 q_3}^{*}
\Bigl[
C_1(\mu_b)\, Q_1^{q_1 q_2 q_3}
+ C_2(\mu_b)\, Q_2^{q_1 q_2 q_3}
\Bigr]
\nonumber \\
&\quad
- V_{tb}\, V_{t q_3}^{*}
\sum_{k=3}^{6} C_k(\mu_b)
\sum_{q = u,d,s,c}
 Q_k^{qq_3}
\Bigg\}
+ {\rm h.c.}\, ,
\label{eq:Heff-SM}
\end{align}
where $V_{qq^\prime}$ are the elements of the Cabibbo-Kobayashi-Maskawa~(CKM) matrix and $G_F$ denotes the Fermi constant.
The $\abs{\Delta B} = 1$ four-quark operators entering \cref{eq:Heff-SM} consist of the current-current operators 
$Q_{1,2}^{q_1 q_2 q_3}$ and of the QCD penguin operators $Q_{3,\ldots,6}^{qq_3}$. These are defined as following~\footnote{Note the different notation for the current-current operators. In our convention, the colour-singlet operator is denoted by $Q_1$ and the colour-rearranged by $Q_2$, just reversed to the notation used in Ref.~\cite{Buchalla:1995vs}.}
\begin{equation}
Q_1^{q_1 q_2 q_3} 
 =   
(\bar q_1^i \, \gamma_\mu P_L \, b^i)
(\bar q_3^j \, \gamma^\mu P_L \, q_2^j)\,,
\qquad 
Q_2^{q_1 q_2 q_3} 
 =  
(\bar q_1^i  \, \gamma_\mu P_L \, b^j)
(\bar{q}_3^j \, \gamma^\mu P_L \, q_2^i)\,,
\label{eq:O12}
\end{equation}
\begin{align}
Q_3^{qq_3} 
& 
= (\bar q_3^i \, \gamma_\mu P_L \, b^i) ( \bar q^j \, \gamma^\mu P_L \, q^j)
\,, \qquad 
Q_4^{qq_3} = (\bar q_3^i \, \gamma_\mu P_L \, b^j)  (\bar q^j \, \gamma^\mu P_L \, q^i)\,, 
\label{eq:O34}
\\[2mm]
Q_5^{qq_3} 
& 
=  (\bar q_3^i \, \gamma_\mu P_L \, b^i) (\bar q^j \, \gamma^\mu P_R \, q^j)\,, 
\qquad
Q_6^{qq_3} = (\bar q_3^i \, \gamma_\mu P_L \, b^j) 
(\bar q^j \, \gamma^\mu P_R \, q^i)\,.
\label{eq:O56}
\end{align}
Here, $i,j = 1, 2, 3,$ label the SU(3)$_c$ indices for fields in the fundamental representation and the projectors are defined as $P_{L,R} = (1 \mp \gamma_5)/2$.\\
In \cref{eq:Heff-SM}, the Wilson coefficients $C_k (\mu_b)$, with $k = 1, \ldots,6$,
are determined at the renormalisation scale $\mu_b \sim m_b$.
They are currently known up to NNLO-QCD accuracy~\cite{Gorbahn:2004my}.
In the SM these coefficients are real-valued, whereas in generic extensions with heavy new physics~(NP), integrated out at some large mass scale $\Lambda_{\mathrm{NP}}$, they typically become complex-valued.\\
Using the optical theorem, the total decay width of a $B$ meson can be computed as
\begin{equation}
    \Gamma (B) = \frac{1}{2 m_B}{\rm Im} \langle B| i \int d^4 x \, T\, \{ {\cal H}_{\rm eff}(x), {\cal H}_{\rm eff}(0)\} | B \rangle \,,
    \label{eq:GammaB}
\end{equation}
in terms of the imaginary part of the forward matrix element of the time-ordered product of the double insertion of the effective Hamiltonian between hadronic $B$-meson states. Taking into account that the $b$ quark is heavy, that is  $m_b \gg \Lambda_{\rm QCD}$, where $\Lambda_{\rm QCD}$ denotes a typical non-perturbative hadronic scale of a few hundreds MeV, 
its momentum can be expressed as 
\begin{equation}
p_b^\mu = m_b v^\mu + i D^\mu \,.
\label{eq:pb_dec}
\end{equation}
Here, $v^\mu$ is the $B$-meson four-velocity, while $D^\mu$ denotes the covariant derivative with respect to the background gluon field, i.e.\ $D_\mu = \partial_\mu - i g_s t^a A^a_\mu$, where $g_s$ is the strong coupling constant, $t^a$ are the SU(3)$_c$ generators in the fundamental representation, and $A^a_\mu$  the gluon field. Owing to \cref{eq:pb_dec}, the $b$-quark momentum is thus decomposed into a leading term proportional to $m_b$ and a residual component associated with soft-gluon interactions within the hadronic state. \\
Using the above relation, the non-local operator in \cref{eq:GammaB} can be computed in terms of a systematic expansion in inverse powers of the heavy $b$-quark mass. The resulting expression, known as the HQE reads~\footnote{This formalism extends to inclusive decays of a hadron $H_Q$ containing a heavy quark $Q$. This includes $b$ baryons and, assuming that the charm quark is heavy enough, charmed hadrons, see e.g.\ Refs.~\cite{King:2021xqp, Gratrex:2022xpm}.}
\begin{equation}
\Gamma(B) = 
\Gamma_3  +
\Gamma_5 \frac{\langle { O}_5 \rangle}{m_b^2} + 
\Gamma_6 \frac{\langle { O}_6 \rangle}{m_b^3} + \ldots  
 + 16 \pi^2 
\left( 
  \tilde{\Gamma}_6 \frac{\langle \tilde{{O}}_6 \rangle}{m_b^3} 
+ \tilde{\Gamma}_7 \frac{\langle \tilde{{O}}_7 \rangle}{m_b^4} + \ldots
\right)\,,
\label{eq:HQE}
\end{equation}
where $\Gamma_d$, with $d = 3,5, \ldots$, are short-distance functions calculable in perturbative QCD:
\begin{equation}
    \Gamma_d = \Gamma_d^{(0)} 
    + \frac{\alpha_s (m_b)}{4 \pi} \Gamma_d^{(1)}  
    + \left(\frac{\alpha_s (m_b)}{4 \pi} \right)^2 \Gamma_d^{(2)}  
    + \ldots \, ,
\end{equation}
and $\langle { O}_d \rangle \equiv
\langle B | { O}_d |B \rangle/(2 m_{B})$ denote 
non-perturbative matrix elements of local $\Delta B = 0$ operators of increasing mass-dimension $d$. A schematic representation of the HQE is shown in \cref{fig:HQEDiags}.
\\
The leading term on the r.h.s.\ of \cref{eq:HQE} corresponds to the inclusive decay width of a free $b$ quark. This contribution is currently known up to NNLO in QCD: the $\alpha_s^2$-corrections arising from the double insertion of the current–current operators in \cref{eq:Heff-SM} were computed recently in Ref.~\cite{Egner:2024azu}.~\footnote{See Ref.~\cite{Egner:2024lay} for a study of the important phenomenological consequences of these corrections.}
Moreover, the NLO corrections due to the insertion of the current-current operators were obtained in Refs.~\cite{Bagan:1994zd,Bagan:1995yf,Krinner:2013cja}, while the contribution of the QCD-penguin operators was determined in \cite{Lenz:1997aa,Lenz:1998qp,Krinner:2013cja}.~\footnote{The contribution of the chromomagnetic  operator $Q_{8}$ has been determined in Refs.~\cite{Greub:2000an,Greub:2000sy}; these corrections will not be discussed further in our work.}
Note that for semileptonic decays, even N3LO-QCD corrections are known~\cite{Fael:2020tow,Czakon:2021ybq,Fael:2023tcv}.\\
Turning to the power-suppressed terms, the contribution at order $1/m_b^2$ was computed at LO-QCD in Refs.~\cite{Bigi:1992su,Blok:1992hw,Blok:1992he}, and recently NLO-QCD corrections were determined in Refs.~\cite{Mannel:2023zei,Mannel:2024uar,Mannel:2025fvj}.
At order $1/m_b^3$, for non-leptonic $b$-quark decays, only LO results are known~\cite{Lenz:2020oce,Mannel:2020fts,Moreno:2020rmk}. We emphasise that, prior to the present work, the contributions of the QCD-penguin operators in \cref{eq:Heff-SM} at dimensions-five and six had not been determined.
As for the semileptonic decays of heavy quarks, power-suppressed contributions have been computed at NLO-QCD both at order $1/m_b^2$ in Refs.~\cite{Alberti:2013kxa,Mannel:2014xza,Mannel:2015jka} and at order $1/m_b^3$ in Refs.~\cite{Mannel:2019qel,Mannel:2021zzr,Moreno:2022goo}. Starting at dimension-six, also four-quark operator contributions enter \cref{eq:HQE}, here denoted with a tilde. At order $1/m_b^3$, these contributions are known to NLO-QCD accuracy~\cite{Beneke:2002rj,Franco:2002fc}, while at order $1/m_b^4$ the accuracy limits to LO-QCD~\cite{Gabbiani:2003pq,Gabbiani:2004tp,Lenz:2013aua}.\\
The matrix elements of the two-quark operators,
$\langle {O}_{5,6} \rangle$ are typically determined from experimental fits to moments of inclusive, semileptonic $B$-meson decays, see e.g.\ Ref.~\cite{Finauri:2023kte}.
As for the spectator-quark effects, the matrix element of the $\Delta B= 0$ four-quark operators, $\langle \tilde{{O}}_6 \rangle$, have been determined within HQET sum rules in Refs.~\cite{Kirk:2017juj,King:2021jsq,Black:2024bus}.
\\For phenomenological applications, one may either compare the measured lifetime of a $B$ meson, $\tau^{\rm Exp.} (B) $, with the HQE prediction for the total decay rate
$\Gamma^{\rm HQE}(B) $ and test to what extent
\begin{equation}
\tau^{\rm Exp.} (B) = 
\frac{1}{\Gamma^{\rm HQE}(B)}
\end{equation}
holds, or instead study lifetime ratios.
In particular, for $\tau(B^0_s)/\tau(B^0_d)$, one obtains
\begin{eqnarray}
\frac{\tau (B^0_s)}{\tau (B^0_d)}
& = & 
\frac{\Gamma (B^0_d)}{\Gamma (B^0_s)}
=
1 + \tau (B^0_s)^{\rm Exp.} \cdot \left[ \Gamma (B^0_d) -  \Gamma (B^0_s) \right]^{\rm HQE} 
 \label{eq:ratio}
\\
& = &  
1 +   \tau (B_s)^{\rm Exp.} \cdot \left[ 
0
+
\Gamma_5 \frac{\langle { O}_5 \rangle_{B^0_d} - 
            \langle { O}_5 \rangle_{B^0_s}}{m_b^2} 
+    
\Gamma_6 \frac{\langle { O}_6 \rangle_{B^0_d} - 
            \langle { O}_6 \rangle_{B^0_s}}{m_b^3}
            + \ldots
            \right.
\nonumber
\\
&&
            \left.
+ 16 \pi^2 
\left( 
  \frac{ \left( \tilde{\Gamma}_6 \langle \tilde{{O}}_6 \rangle \right)_{B^0_d} -
  \left( \tilde{\Gamma}_6 \langle \tilde{{O}}_6 \rangle \right)_{B^0_s}}{m_b^3} 
+ 
\frac{ \left( \tilde{\Gamma}_7 \langle \tilde{{O}}_7 \rangle \right)_{B^0_d} -
  \left( \tilde{\Gamma}_7 \langle \tilde{{O}}_7 \rangle \right)_{B^0_s}}{m_b^4} 
+ \ldots
\right)
\right]^{\rm HQE} \, .
\nonumber
\end{eqnarray}
In lifetime ratios, the free-quark decay $\Gamma_3$ cancels exactly and here this is explicitly indicated by the ``0'' in the second line of \cref{eq:ratio}.
The contribution of the chromomagnetic operator at dimension-five and of the Darwin operator at dimension-six cancel up to SU(3)$_F$-breaking corrections. These corrections can be sizable and therefore cannot, in general, be neglected, see e.g.\ Ref.~\cite{Lenz:2022rbq}.
In contrast, for the ratio $\tau(B^+)/\tau(B_d^0)$ these contributions cancel almost completely as a consequence of isospin symmetry.
As for the $16\pi^2$ enhanced spectator-quark effects, here both the Wilson coefficients $\tilde{\Gamma}_{d=6,7, \ldots}$ and the corresponding hadronic matrix elements differ for different hadrons.
Their contribution to the lifetime ratio $\tau(B^0_s)/\tau(B^0_d)$ is relatively small~\cite{Lenz:2022rbq}, while it is large, i.e.\ ${\cal O} (10\%$), for the $\tau(B^+)/\tau(B^0_d)$~
\cite{Lenz:2022rbq} due to the large Pauli-interference contributions.~\footnote{For comparison, these effects are even more pronounced in the charm sector, where they exceed $100\%$ in the lifetime ratio $\tau(D^+)/\tau(D^0)$, see Ref.~\cite{King:2021xqp}.}\\
To account for possible BSM contributions to non-leptonic, tree-level $b$-quark decays, the effective Hamiltonian in \cref{eq:Heff-SM} must be extended to include the most general $\abs{\Delta B} = 1$ four-quark operator basis. This extension is discussed in the next section.
\subsection{Generalised effective Hamiltonian}
\label{subsec:Heff}
The most general model-independent effective Hamiltonian describing non-leptonic decays $b\to q_1 \bar q_2 q_3$, with $q_1, q_2 = u,c,$ and $q_3 = d,s,$ including all possible Dirac structures, both those that appear in the SM and additional ones beyond it, is given by~\cite{Ciuchini:1997bw,Buras:2000if}
\begin{equation}
{\cal H}_{\rm eff}
= 
\frac{4 G_F}{\sqrt 2} V_{q_1 b} V_{q_2 q_3}^*
\sum_{i = 1}^{10} 
\Bigl[
{\cal C}_i (\mu_b) {\cal Q}_i^{q_1q_2q_3} + {\cal C}_i^{\prime} (\mu_b) {\cal Q}_i^{q_1q_2q_3 \prime} 
\Bigr] 
+ {\rm h.c.}\,.
\label{eq:Heff-NP}    
\end{equation}
Here, the four-quark operators ${\cal Q}_i^{q_1q_2q_3}$ are defined as
\begin{align}
& {\cal Q}_1^{q_1q_2q_3} = (\bar q_1^i \gamma_\mu P_L \, b^i) (\bar q_3^j \gamma^\mu P_L \, q_2^j)\,,
\quad &
& {\cal Q}_2^{q_1q_2q_3} = (\bar q_1^i \gamma_\mu P_L \, b^j) (\bar q_3^j \gamma^\mu P_L \, q_2^i)\,,
\label{eq:Q1-Q2}
\\[2mm]
& {\cal Q}_3^{q_1q_2q_3} = (\bar q_1^i \gamma_\mu P_R \, b^i) (\bar q_3^j \gamma^\mu P_L \, q_2^j)\,,
\quad &
& {\cal Q}_4^{q_1q_2q_3} = (\bar q_1^i \gamma_\mu P_R \, b^j) (\bar q_3^j \gamma^\mu P_L \, q_2^i)\,,
\label{eq:Q3-Q4}
\\[2mm]
& {\cal Q}_5^{q_1q_2q_3} = (\bar q_1^i P_L \, b^i) (\bar q_3^j P_R \, q_2^j)\,,
\quad &
& {\cal Q}_6^{q_1q_2q_3} = (\bar q_1^i P_L \, b^j) (\bar q_3^j P_R \, q_2^i)\,,
\label{eq:Q5-Q6}
\\[2mm]
& {\cal Q}_7^{q_1q_2q_3} = (\bar q_1^i P_R \, b^i) (\bar q_3^j P_R \, q_2^j)\,, 
\quad &
& {\cal Q}_8^{q_1q_2q_3} = (\bar q_1^i P_R \, b^j) (\bar q_3^j P_R \, q_2^i)\,,
\label{eq:Q7-Q8}
\\[2mm]
& {\cal Q}_9^{q_1q_2q_3} = (\bar q_1^i \sigma_{\mu\nu} P_R\, b^i)(\bar q_3^j \sigma^{\mu\nu} P_R\, q_2^j)\,,
\quad &
& {\cal Q}_{10}^{q_1q_2q_3} = (\bar q_1^i \sigma_{\mu\nu} P_R\, b^j)(\bar q_3^j \sigma^{\mu\nu} P_R\, q_2^i)\,,
\label{eq:Q9-Q10}
\end{align}
where we use the convention $\sigma^{\mu\nu} = (i/2) \left[\gamma^\mu, \gamma^\nu\right]$, and the primed operators ${\cal Q}_i^{q_1q_2q_3 \prime}$ in \cref{eq:Heff-NP} are obtained from those in \cref{eq:Q1-Q2,eq:Q3-Q4,eq:Q5-Q6,eq:Q7-Q8,eq:Q9-Q10} by flipping the corresponding
chiralities, i.e.\ 
\begin{equation}
{\cal Q}_i^{q_1q_2q_3 \prime} = {\cal Q}_i^{q_1q_2q_3}\big|_{P_L \leftrightarrow P_R}\,.
\end{equation}
In \cref{eq:Heff-NP}, the Wilson coefficients ${\cal C}_i^{(\prime)}$ are, in general, complex numbers.  For brevity, we do not explicitly indicate their dependence on the flavour structure $(q_1 q_2 q_3)$.\\
The generalised effective Hamiltonian in \cref{eq:Heff-NP} includes the SM one in \cref{eq:Heff-SM}.
In particular, the operators ${\cal Q}_1^{q_1 q_2 q_3}$ and ${\cal Q}_2^{q_1 q_2 q_3}$ in \cref{eq:Q1-Q2} coincide with the SM current-current operators $Q_1^{q_1 q_2 q_3}$ and $Q_2^{q_1 q_2 q_3}$, namely
\begin{equation}
    Q_1^{q_1q_2q_3} = {\cal Q}_1^{q_1q_2q_3} \,, \qquad Q_2^{q_1q_2q_3} = {\cal Q}_2^{q_1q_2q_3}\,.
    \label{eq:Q1_Q1cal}
\end{equation}
Furthermore, Fierz transformations relate the operators ${\cal Q}_{1,2}^{q q q_3}$ and ${\cal Q}_{5,6}^{q q q_3}$ to the SM QCD-penguin operators $Q_3^{qq_3}, \ldots, Q_6^{qq_3}$, for $q = u,c$.
Specifically:
\begin{align}
     Q_3^{qq_3} &= {\phantom{-2}}{\cal Q}_2^{qqq_3} \,, 
    \, 
  & Q_4^{qq_3} &= {\phantom{-2}}{\cal Q}_1^{qqq_3} \,,
  \label{eq:O3_O4}\\[2mm]
    Q_5^{qq_3} &= -2  {\cal Q}_6^{qqq_3} \,, 
    \,
   & Q_6^{qq_3} &=-2 {\cal Q}_5^{qqq_3} \,.
   \label{eq:O5_O6}
\end{align}

Correspondingly, the Wilson coefficients $\mathcal{C}_{i}^{(\prime)}$ can be decomposed into a real SM contribution and a purely BSM contribution, which, as already stated, is in general complex.   
We may thus write $\mathcal{C}_{i}^{(\prime)} = \mathcal{C}_{i,{\mathrm{SM}}}^{(\prime)} + \mathcal{C}_{i,\mathrm{NP}}^{(\prime)}$, where, for instance, ${\cal C}_{1, {\rm SM}} = C_1$ and  ${\cal C}^\prime_{1, {\rm SM}} = 0$.
It follows that 
the double insertion of $\mathcal{H}_{\mathrm{eff}}$ into \cref{eq:GammaB} accounts for all possible contributions to the decay rate $\Gamma$: purely SM, purely BSM, and the mixed SM--BSM interference terms.
We therefore choose \cref{eq:Heff-NP} as the starting point for our calculation and all results presented below refer to this basis of operators.\\
We emphasise that, so far, the contributions of the generalised Hamiltonian in \cref{eq:Heff-NP} to the HQE have only been partially determined. Dimension-six four-quark operator contributions, $\tilde \Gamma_6^{(0)}$, were computed for the $b \to c \bar c s$ transition in Refs.~\cite{Jager:2017gal,Jager:2019bgk} and for $b \to c \bar u d$ in Ref.~\cite{Lenz:2022pgw}. More recently, the leading-power contributions for the $b \to c \bar u d$ channel were derived in Ref.~\cite{Meiser:2024zea} in LO-QCD. In addition, hadronic matrix elements of dimension-six four-quark operators with generic BSM Dirac structures have been determined in the HQET limit using three-loop sum rules in Ref.~\cite{Black:2024bus}.\\
In this work, we compute the remaining missing contributions to the HQE, namely the matching coefficients of the two-quark operators up to dimension-six.
\subsection{Outline of the calculation}
\label{Calsteps}
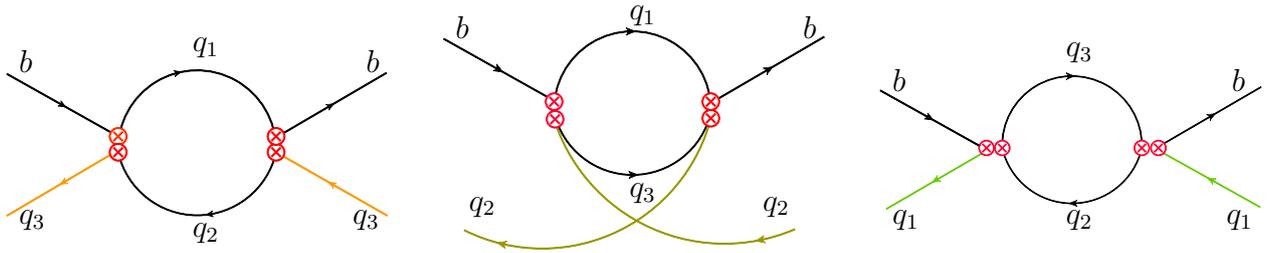
\begin{figure}[t!]
  \centering
  \setlength{\unitlength}{1pt}
  \begin{tabular}{ccccc}
    \begin{picture}(150,120)
      \put(0,32){\includegraphics[width=0.30\textwidth]{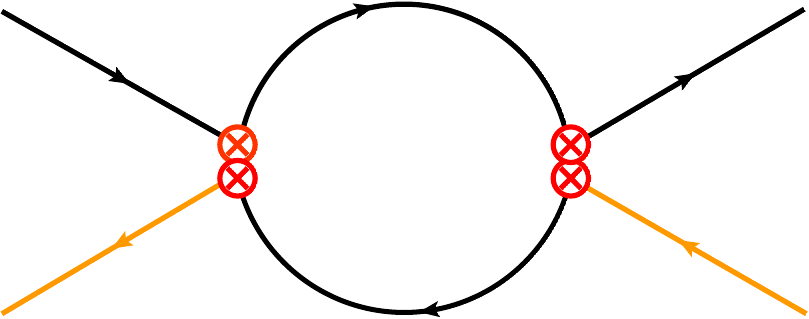}}
      \put(70,95){$q_1$}
      \put(70,25){$q_2$}
      \put(5,30){$q_3$}
      \put(130,30){$q_3$}
      \put(5,87){$b$}
     \put(135,87){$b$}
    \end{picture}
    \hspace{0.2cm}
    \begin{picture}(150,120)
      \put(0,20){\includegraphics[width=0.30\textwidth]{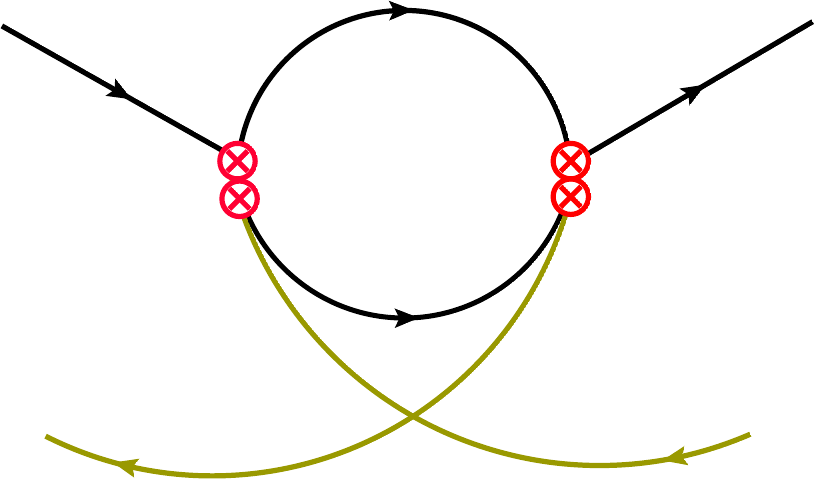}}
      \put(70,107){$q_1$}
      \put(70,40){$q_3$}
      \put(10,35){$q_2$}
      \put(120,35){$q_2$}
      \put(5,100){$b$}
     \put(135,100){$b$}
    \end{picture}
\hspace{0.2cm}
    \begin{picture}(150,120)
      \put(0,35){\includegraphics[width=0.30\textwidth]{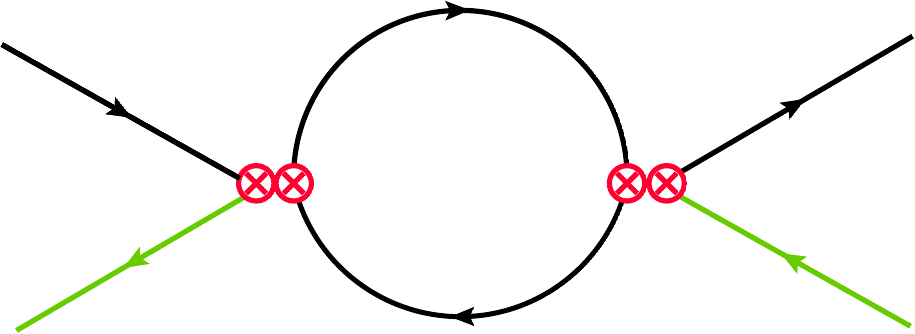}}
      \put(70,93){$q_3$}
      \put(70,30){$q_2$}
      \put(5,30){$q_1$}
      \put(130,30){$q_1$}
      \put(5,80){$b$}
     \put(132,80){$b$}
    \end{picture}
  \end{tabular}
\caption{Diagrammatic representation of the $\Delta B  = 0$ four-quark operator contributions in the HQE from the $b\to q_1 \bar q_2 q_3$ transition: WE (left), PI (middle), and WA (right).}
\label{fig:4QDiags}
\end{figure}

The main scope of this work is to compute the matching coefficients of the two-quark operators in the HQE, up to dimension-six, namely $\Gamma_3^{(0)}$, $\Gamma_5^{(0)}$, and $\Gamma_6^{(0)}$ in \cref{eq:HQE}, arising from the generalised effective Hamiltonian in~\cref{eq:Heff-NP}. A diagrammatic representation of these contributions is given by the left and middle panels of \cref{fig:HQEDiags}. \\
As discussed in \cref{sec:Intro}, the leading term  $\Gamma_3$ describes the contribution of the free $b$-quark decay, with the corresponding dimension-three matrix element parametrised only by the $B$-meson mass $m_B$. At subleading order, the terms $\Gamma_5$ and $\Gamma_6$ account for power corrections from dimension-five and dimension-six operators: the kinetic and chromomagnetic operators at dimension-five, and the Darwin and spin-orbit operators at dimension-six.
Their matrix elements are further parametrised by the non-perturbative inputs $\mu_\pi^2$, $\mu_G^2$, $\rho_D^3$, and $\rho_{LS}^3$, respectively, which encode the expectation values of the corresponding two-quark operators within the $B$-meson state. These are defined as  
\begin{equation}
\begin{aligned}
2 m_B\,\mu_\pi^2(B) &= -\langle B(p_B)\,|\,\bar b_v (iD_\mu)(iD^\mu)\,b_v\,|\,B(p_B)\rangle \,, \\[4pt]
2 m_B\,\mu_G^2(B) &= \langle B(p_B)\,|\,\bar b_v (iD_\mu)(iD_\nu)(-i\sigma^{\mu\nu})\,b_v\,|\,B(p_B)\rangle \,, \\[6pt]
2 m_B\,\rho_D^3(B) &= \langle B(p_B)\,|\,\bar b_v (iD_\mu)(i v\!\cdot\!D)(iD^\mu)\,b_v\,|\,B(p_B)\rangle \,, \\[6pt]
2 m_B\,\rho_{LS}^3(B) &= \langle B(p_B)\,|\,\bar b_v (iD_\mu)(i v\!\cdot\!D)(iD_\nu)(-i\sigma^{\mu\nu})\,b_v\,|\,B(p_B)\rangle \,,
\end{aligned}
\label{eq:hadronic_parameters}
\end{equation}  
where $b_v$ denotes the rescaled $b$-quark field
\begin{equation}
    b(x) = e^{-i m_b v\cdot x}\,b_v(x) \,,
\end{equation}
with $v^\mu = p_B^\mu/m_B$ being the four-velocity of the $B$ meson.\\
As already mentioned, at dimension-six, the computation of the coefficient of the Darwin operator gives rise IR to divergences.
These originate from the emission of soft gluons from the propagator of one of the light quarks $q = u,d,s$ and signal operator mixing with the corresponding $\Delta B = 0$ four-quark operators with $q$ quarks~\cite{Piscopo:2021ogu}.
In order to regularise these divergences, in our calculation we have considered two different schemes. Specifically we have performed the computation directly in $D = 4$ dimensions, introducing a mass regulator for the light quarks, $m_q$, as well as in dimensional regularisation (DR) with $D=4-2\epsilon$ and vanishing light-quark masses. 
These two methods partly follow the two independent approaches employed in Ref.~\cite{Lenz:2020oce} and Ref.~\cite{Mannel:2020fts} for the computation of the SM coefficient of the Darwin operator for non-leptonic $b$-quark decays. 
For clarity, below we present the calculation performed in four dimensions using a mass regulator, 
while the details of the independent computation performed in DR are outlined in \cref{sec:DR}. 
Both approaches lead to identical expressions for all physical Wilson coefficients, providing a crucial and non-trivial consistency check of our results.\\
The starting point of the calculation is the expansion of the internal $q_1, q_2,$ and $q_3$ quark propagators in the external background gluon field using the Fock-Schwinger (FS) gauge, see e.g.\ Ref.~\cite{Novikov:1984ecy}. 
The corresponding expression up to a single covariant derivative of the gluon field strength tensor $G_{\mu \nu} = -i  \left[iD_\mu, iD_\nu \right]$, which suffices for our computation, reads~\cite{Lenz:2020oce}
\begin{equation}
S_{q_i}(x,0) =  \int \frac{d^4 p}{(2\pi)^4}\, {\cal S}_{q_i}(p)  \, e^{-i p \cdot x} \,,
\label{eq:Sx0}
\end{equation}\\
with
\begin{align}
{\cal S}_{q_i}(p) & = \frac{\slashed p + m_i}{p^2 - m_i^2 + i \varepsilon } -\frac{m_i }{2} \frac{G_{\rho \mu } \,\sigma^{\rho \mu}}{(p^2 - m_i^2 + i \varepsilon)^2}  +  \frac{\tilde G_{\sigma \eta} \ p^\sigma \gamma^\eta \gamma^5}{(p^2 - m_i^2 + i \varepsilon)^2} - \frac23 \frac{p^\alpha D_\alpha  G_{\rho \mu}}{(p^2 - m_i^2 + i \varepsilon)^3} \gamma^\mu p^\rho 
\nonumber \\[3mm]
&+ \frac23 \frac{D^\alpha G_{\alpha \mu}}{(p^2 - m_i^2 + i \varepsilon)^3} \Big[ \gamma^\mu  (p^2 - m_i^2) - p^\mu (\slashed p + 2 m_i) \Big]
 + 2 i \frac{D_\alpha \tilde G_{\tau \eta}}{(p^2 -m_i^2 + i \varepsilon)^3} p^\alpha p^\tau \gamma^\eta \gamma^5 
\nonumber \\[3mm]
& + \frac23 m_i \frac{D_\alpha G_{\rho \mu}}{(p^2 -m_i^2 + i \varepsilon)^3} \Big( p^\alpha \gamma^\rho \gamma^\mu - p^\rho \gamma^\mu \gamma^\alpha \Big)
+ \, \ldots \,,
\label{eq:S1p}
\end{align}
where the ellipses denote higher-dimensional terms; the dual field tensor is $\tilde G_{\mu \nu} = (1/2) \epsilon_{\mu \nu \rho \sigma} G^{\rho \sigma}$ with the Levi-Civita tensor $\epsilon_{\mu \nu \rho \sigma}$; and $ D_\rho G_{\mu \nu} = -\left[ iD_\rho,  \left[iD_\mu, iD_\nu \right] \right]$.\footnote{Note that the FS gauge breaks translation symmetry, so that $S_{q_i}(x,0) \neq S_{q_i}(0,x)$. For the explicit form of $S_{q_i}(0,x)$ we refer to Ref.~\cite{Lenz:2020oce}. Here we only display $S_{q_i}(x,0)$ for brevity.} \\
The first term on the r.h.s.\ of \cref{eq:S1p} corresponds to the free-quark propagator, whose contribution is sufficient for the calculation of the leading dimension-three term in the HQE. To account for power corrections at dimension-five, the second and third terms on r.h.s.\ of \cref{eq:S1p} must also be included. The remaining terms contribute only at dimension-six.  \\
Using $b(x) = e^{-i p_b \cdot x} \, b(0)$, where $p_b$ is the external $b$-quark momentum, and integrating over the coordinate $x$ and the loop momenta, yields two-loop integrals of the form 
\begin{equation}
\hspace*{-3.5mm}
 {\cal I}_{n_1 n_2 n_3}
\equiv 
\int \frac{d^4 l_1}{(2 \pi)^4} \, \int \frac{d^4 l_2}{(2 \pi)^4}  \frac{ f(p_b, l_1, l_2) }{ \big[ l_1^2 - m_{1}^2 + i \varepsilon \big]^{n_1} \big[ l_2^2 - m_{2}^2 + i \varepsilon \big]^{n_2} \big[ (l_1 + l_2 -p_b)^2 - m_{3}^2 + i \varepsilon \big]^{n_3}} \,,
\label{eq:MI-def}
\end{equation}
where $n_i \in {\mathbb N}_0$, and $f(p_b, l_1, l_2)$ depends on all scalar products of its arguments. From \cref{eq:S1p}, in our calculation integrals with $n_i$ up to 3 appear.
These integrals, being free of ultraviolet divergences, are evaluated directly in four dimensions and reduced to master integrals with known imaginary parts, see Ref.~\cite{Remiddi:2016gno,Mannel:2020fts}. 
Dirac algebra simplifications is performed using \texttt{FeynCalc}~\cite{Mertig:1990an,Shtabovenko:2016sxi}, and integration-by-parts (IBP) reductions is carried out with \texttt{LiteRed}~\cite{Lee:2012cn}.\\
After performing the loop integration and computing the corresponding imaginary part, the two-quark operator matrix elements can be expressed in the general form
\begin{equation}
\begin{aligned}
 \mathcal{G}(p_b) \ 
 \langle B(p_B)\,|\,\bar b_v(0) \, 
 \mathcal{M}^{\mu_1 \dots \mu_3} \,
 (iD_{\mu_1}) \dots (iD_{\mu_3}) \,
 b_v(0) \,|\, B(p_B) \rangle \,,
\end{aligned}
\label{eq:MEGeneral}
\end{equation}
where $\mathcal{G}(p_b)$ encodes the dependence on the internal quark masses and the external momentum, and $\mathcal{M}_{\mu_1 \dots \mu_3}$ specifies the Dirac structure.
At order $\mathcal{O}(1/m_b^3)$, contributions with up to three covariant derivatives are retained.
Expanding ${\cal G}(p_b)$ in powers of $D^\mu/m_b$, taking into account \cref{eq:pb_dec}, leads to a tower of local operators containing increasing numbers of covariant derivatives.
The Lorentz indices of multiple derivatives originating from factors $p_b$ are symmetrised according to
\begin{equation}
p_b^{\mu_1} p_b^{\mu_2} \dots p_b^{\mu_n}
= \frac{1}{n!} \sum_{\sigma \in S_n}
p_b^{\sigma(\mu_1)} p_b^{\sigma(\mu_2)} \dots p_b^{\sigma(\mu_n)} \,,
\end{equation}
where $S_n$ denotes the set of all permutations of $n$ indices.
The matrix elements of the resulting local operators are then mapped onto the HQE basis using the trace formalism provided in Refs.~\cite{Dassinger:2006md, Mannel:2023yqf}, leading to the terms of dimension-three, five and six. In the latter two cases, the corresponding contributions are parametrised by the non-perturbative inputs $\mu_\pi^2$, $\mu_G^2$, and $\rho_D^3$, cf.\ \cref{eq:hadronic_parameters}.\footnote{Note that, with our conventions for the two-quark operators defined in \cref{eq:hadronic_parameters}, the contribution of the spin-orbit operator $\rho_{LS}^3$ vanishes identically, see e.g.\ Ref.~\cite{Mannel:2018mqv}.}
\begin{figure}[t!]
  \centering
  \setlength{\unitlength}{1pt}

  \begin{tabular}{ccccc}
    \begin{picture}(150,120)
      \put(0,0){\includegraphics[width=0.27\textwidth]{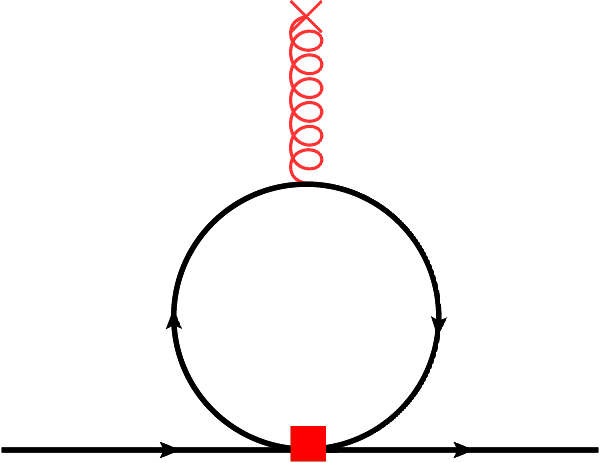}}
      \put(25,35){$q$}
      \put(97,35){$q$}
      \put(10,8){$b$}
      \put(110,8){$b$}
    \end{picture}
     \end{tabular}
     \caption{One soft-gluon correction to the matrix element of the dimension-six $\Delta B = 0$ four-quark operators with light $q$ quarks, leading to operator mixing with the Darwin operator. The inclusion of these contributions ensures the cancellation of the IR divergences appearing in the matching coefficient of the Darwin operator.}
\label{fig:MatrixMix}
    \end{figure}
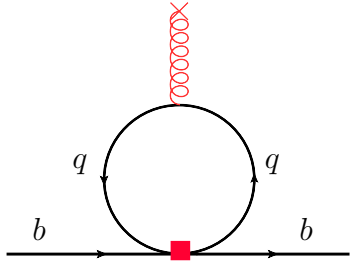
\\While the matching coefficients at dimensions-three and -five can be now obtained directly, at dimension-six the calculation of the Darwin-operator contribution requires additional care.
As already discussed at the beginning of the section, at this order, the expansion of the propagator in \cref{eq:S1p} for $q_i = q = u,d,s$ leads to IR divergences.
These are regularised by retaining a finite light-quark mass $m_q$, resulting in logarithmic terms of the type $\log(m_q^2/m_b^2)$.
These logarithms cancel against the renormalised one-loop matrix elements of dimension-six $\Delta B = 0$ four-quark operators with $q$ quarks, see \cref{fig:MatrixMix}, ensuring finite final results as $m_q \to 0$.\\
Specifically, starting from the generalised effective Hamiltonian in \cref{eq:Heff-NP}, the resulting $\Delta B = 0$ four-quark operators $\mathcal{O}_i$ and $\mathcal{O}_i^{\prime}$ are defined as~\cite{Lenz:2022pgw} 
\begin{align}
  \mathcal{O}_1
  & =  
  4\, (\bar b \, \gamma_\mu P_L \, q)\,(\bar{q} \, \gamma^\mu P_L \, b) \,,
  &
  \mathcal{O}_2  
  & =  
  4 \, (\bar b P_L \, q)\,(\bar{q} P_R \, b) \,,
  \label{eq:O1-O2-HQET} 
  \\[1mm]
  \mathcal{O}_3  
  & =  
  4 \,(\bar b \, \gamma_\mu P_L \, t^a q)\,(\bar{q}\,\gamma^\mu P_L\,t^a b) \,,
  &
  \mathcal{O}_4 
  & =  
  4\,(\bar b P_L \, t^a q)\,(\bar{q} P_R \, t^a b) \,,
  \label{eq:O3-O4-HQET} 
  \\[1mm]
  \mathcal{O}_5  
  & =  
  4\,(\bar b \, \gamma_\mu P_L \, q)\, (\bar{q} \, \gamma^\mu P_R \, b) \,,
  &
  \mathcal{O}_6  
  & =  
  4 \, (\bar b P_L \, q) \, (\bar{q} P_L \, b) \,,
  \label{eq:O5-O6-HQET} 
  \\[1mm]
  \mathcal{O}_7  
  & =  
  4\,(\bar b \, \gamma_\mu P_L \, t^a q) (\bar{q}\,\gamma^\mu P_R\,t^a b) \,,
  &
  \mathcal{O}_8 
  & =  
  4 \, (\bar b P_L \, t^a q) \,
  (\bar{q}P_L \, t^a b) \,,
  \label{eq:O7-O8-HQET}
\end{align}
while the primed operators $\mathcal{O}^{\prime}_i$ are obtained by replacing $P_L \leftrightarrow P_R$ in \cref{eq:O1-O2-HQET,eq:O3-O4-HQET,eq:O5-O6-HQET,eq:O7-O8-HQET}.
Furthermore, one can write additional $\Delta B = 0$ operators with tensor structures. However, they are, in fact, not independent at the dimension-six level. Using the $b$-quark equations of motion, one can express them, up to power corrections, as linear combinations of operators with the remaining Dirac structures~\cite{Lenz:2022pgw},
\begin{align}
	(\bar b \sigma^{\rho \tau} q)(\bar q \sigma_{\rho \tau} b) = 
	- 2 \left[(\bar b  q)(\bar q  b) 
	- (\bar b \gamma^\rho q) (\bar q \gamma_\rho b) 
	+ (\bar b \gamma_5 q)(\bar q \gamma_5 b)  
	+ (\bar b \gamma^\rho \gamma_5 q)(\bar q \gamma_\rho \gamma_5 b)\right] + {\cal O}\left(\frac{1}{m_b}\right)\,.
	\nonumber\\
\label{eq:tensor-EOM-relation}
\end{align}
The renormalised dimension-six matrix element of the operator on the l.h.s.\ of the above equation vanishes.
The corresponding matrix elements of the individual terms on the r.h.s.\ are non-zero; however, their sum also vanishes, providing an additional cross-check of the relation in \cref{eq:tensor-EOM-relation}.\\
We note that for the calculation of the coefficients relevant to the $b \to c\bar u d$ and $b \to c\bar c s$ transitions,
the cancellation of all IR divergences is ensured by including the results for the
Pauli interference~(PI) and weak exchange (WE) contributions, whose diagrams are shown in the left and middle panels of \cref{fig:4QDiags}. These contributions were already calculated in
Ref.~\cite{Lenz:2022pgw} for the generalised Hamiltonian in \cref{eq:Heff-NP} and we use the expressions presented there.
For the channels $b \to u\bar c s$ and
$b \to u\bar u d$, however, these contributions alone are not sufficient:
the divergent logarithms originating from the emission of a soft gluon from the $q_1$-quark propagator cancel only after also accounting for the contribution of
weak annihilation (WA) diagrams, schematically shown in the right panel of \cref{fig:4QDiags}. We have therefore computed the corresponding WA contributions for the Hamiltonian in \cref{eq:Heff-NP}, and we present the resulting expressions in \cref{sec:WA}.\\
The one-loop matrix elements of the $\Delta B = 0$ four-quark operators contain both ultraviolet (UV) and IR divergences. To regularise the former we use DR and subtract the associated $1/\epsilon$ poles in the $\overline{\text{MS}}$ scheme. 
For the latter we must choose the same regularisation scheme applied to the computation
of the matching coefficients of the two-quark operator contributions discussed above, hence we keep the mass of the light-quark
running in the loop finite.
The treatment of the divergences follows the same cancellation pattern discussed in Ref.~\cite{Lenz:2020oce}.\\
Under renormalisation, only the matrix elements of  the operators ${\cal O}_{1-4}$, which also arise in the SM, and of their chiral counterparts ${\cal O}'_{1-4}$ are not zero when inserted into \cref{fig:MatrixMix}. 
The corresponding results for the renormalised matrix elements are thus identical to the SM ones and, at the renormalisation scale $\mu = m_b$, take the form
\begin{equation}
\begin{aligned}
\langle {{\cal O}}_{1,2} \rangle^{\text{ren}} &= \langle {\cal O}'_{1,2}\rangle^{\text{ren}} = 
 \frac{a_{1,2}}{12 \pi^2}  
 \left[ 
 \log \!\left( \frac{m_b^2}{m_q^2}\right) + b_{1,2} 
 \right] 
 \langle {{\cal O}}_{\rho_D} \rangle 
 + {\cal{O}}\!\left(\frac{1}{m_b}\right) \,, \\[2mm]
 \langle {\cal O}^{(\prime)}_{3,4} \rangle^{\text{ren}} &= 
 -\frac{1}{2N_c} 
 \langle {\cal O}^{(\prime)}_{1,2}  \rangle^{\text{ren}} \,,
\end{aligned}
\label{eq:crD-finite}
\end{equation}
where ${\cal O}_{\rho_D} = \bar b_v (i D_\mu)(i v\!\cdot\! D)(i D^\mu)b_v$ denotes the Darwin operator. 
The coefficients $a_i$ and $b_i$ coincide with those obtained in the SM, with $a_1 = 2$, $a_2 = -1$, $b_1 = - 1$, and $b_2 = 0$~\cite{Piscopo:2021ogu}.
We stress that the numerical value of these constants is not universal but depends on the chosen operator basis.
For instance, in Ref.~\cite{Lenz:2020oce}, the $\Delta B = 0$ basis is defined in terms of colour-rearranged operators rather than colour-octet ones used here, resulting in different constants.\\
Soft-gluon corrections to the remaining operators ${\cal O}_{5-8}$ and ${\cal O}'_{5-8}$ vanish identically and therefore do not contribute at this order. Note that, although their renormalised one-loop matrix elements vanish, these operators still enter at the tree-level in the WA contribution, see \cref{sec:WA}. \\
Upon cancellation of all IR divergences, we arrive at the final expressions for the coefficient of the Darwin operator, thus completing the determination of the two-quark operator contributions to the HQE up to dimension-six. Our results are discussed in the next section.
\section{Analytic Results} 
\label{Sec:Results}
In this section we present our results for the matching coefficients to the two-quark operators in the HQE of a $B$ meson,\footnote{We emphasise that the results derived here extend straightforwardly to the HQE for $b$-baryons.} originating from the generalised effective Hamiltonian in \cref{eq:Heff-NP}.
Up to subleading ${\cal O}(1/m_b^4)$ terms, the contribution to tree-level, non-leptonic $b \to q_1 \bar q_2 q_3$ decays can be expressed compactly as\footnote{For brevity, we omit to add a suffix $2q$; unless stated otherwise, it should be understood that only the contribution of two-quark operators in the HQE is considered.}
\begin{equation}
\Gamma(b\to q_1 \bar q_2 q_3)
= 
\Gamma_0  |V_{q_1 b}|^2\,|V_{q_2 q_3}|^2 \sum_{i,j=1}^{20}  \, {\cal C}_i \,{\cal C}_j^* \, 
\mathcal{N}_{ij} \mathcal{F}_{i,j}^{(q_1 q_2 q_3)} (\rho) \,,
\label{eq:Gamma_master_compact}
\end{equation}
where the prefactor $\Gamma_0$ is given by
\begin{equation}
\Gamma_0 \equiv
\frac{G_F^2\,m_b^5}{192\,\pi^3}\,,
\label{eq:Gamma0}
\end{equation}
and for simplicity we now denote the contribution from primed operators with the indices $i,j = 11,\dots,20$, i.e.\ ${\cal C}_{11} \equiv {\cal C}^{\prime}_{1}, \dots, {\cal C}_{20} \equiv {\cal C}^{\prime}_{10}$.
The functions $\mathcal{F}_{i,j}^{(q_1 q_2 q_3)}$ in \cref{eq:Gamma_master_compact} depend on the mass ratio $\rho = m_c^2/m_b^2$ and encode the analytic dependence on $\rho$ for each operator pair combination $\mathcal{Q}_{i}^{q_1 q_2 q_3} \mathcal{Q}_j^{\dagger \, q_1 q_2 q_3}$ and at each order in the HQE.
Their power expansion in terms of the non-perturbative parameters $\mu_\pi^2(B), \mu_G^2(B)$ and $ \rho_D^3 (B)$ reads:
\begin{equation}
\begin{aligned}
\mathcal{F}_{i,j}^{(q_1 q_2 q_3)} &\equiv
\left( 1 - \frac{\mu_\pi^2}{2 m_b^2}  \right) P_{i,j}^{(q_1 q_2 q_3)}(\rho)
+ \frac{\mu_G^2}{m_b^2} G_{i,j}^{(q_1  q_2 q_3)}(\rho)
+ \frac{\rho_D^3}{m_b^3} D_{i,j}^{(q_1 q_2 q_3)}(\rho) \,,
\label{eq:Fij}
\end{aligned}
\end{equation}
where we have taken into account that, due to reparametrisation invariance, at order $1/m_b^2$, the coefficient of the kinetic operator $\mu_\pi^2$ coincides with the leading-power contribution up to the constant term $(-1/2)$.
The colour factors ${\cal N}_{ij}$ in \cref{eq:Gamma_master_compact} are given by:
\begin{equation}
\mathcal{N}_{ij}
=
\begin{cases}
    1, &\text{if } j - i \text{ odd}\, \\
    N_c, &\text{if } j - i \text{ even} \,.
\end{cases}
\label{eq:color_matrix}
\end{equation}

Since the coefficient functions ${\cal F}_{i,j}^{(q_1 q_2 q_3)}(\rho)$ obey the following universal relations
\begin{align}
{\cal F}_{i,j}^{(q_1 q_2 q_3)} &= {\cal F}_{j,i}^{(q_1 q_2 q_3)}\,,   \quad
{\cal F}_{i+10,\, j+10}^{(q_1 q_2 q_3)} = {\cal F}_{i,j}^{(q_1 q_2 q_3)}\,, \, \\[2mm]
{\cal F}_{i,j}^{(q_1 q_2 q_3)}(\rho) &= 0 \qquad \text{for } 1 \leq i \leq 10,\ 11 \leq j \leq 20,
\end{align}
for all $i,j$, in the following we present only the independent functions $\mathcal{F}_{i,j} (\rho)$ with $1 \leq i \leq j \leq 10$.\\
We emphasise that our computation has been carried out for four distinct choices of the internal quark masses $m_1, m_2,$ and $m_3$, covering all non-leptonic $b$-quark decay channels.
Specifically:
\begin{itemize}
\item[(i)] $m_1 \neq 0$, $m_2 = m_3 = 0$: These results apply to $b \to c \bar u d$ and $b \to c \bar u s$.
\item[(ii)] $m_1 = m_2 \neq 0$, $ m_3 = 0$: These results apply to $b \to c \bar c s$ and $b \to c \bar c d$.
\item[(iii)] $m_2 \neq 0$, $m_1 = m_3 = 0$: These results apply to $b \to u \bar c s$ and $b \to u \bar c d$.
\item[(iv)] $m_1 = m_2 = m_3 = 0$: These results apply to $b \to u \bar u d$ and $b \to u \bar u s$.
\end{itemize}

Below we present our results only for the CKM dominant modes $b \to c \bar u d$ and $b \to c \bar c s$.
Corresponding expressions for the CKM suppressed channels $b \to u \bar c d$ and $b \to u \bar u d$ are listed in \cref{sec:ucs,sec:uud}, respectively.
For each decay mode, the combinations of SM operators reproduce the known results from Refs.~\cite{Mannel:2020fts,Lenz:2020oce}.
Complete analytic expressions for all eight non-leptonic decay channels are provided in the ancillary file \texttt{fullAmpIncludingCKMSuppressed.m}, together with a minimal usage example in \texttt{minimal\_example.m}.\\
The functions $P_{i,j}^{(q_1q_2q_3)} (\rho)$, $G_{i,j}^{(q_1q_2q_3)} (\rho) $, and $D_{i,j}^{(q_1q_2q_3)} (\rho)$ are the main results of this work.
However, the operator identities in \cref{eq:Q1_Q1cal,eq:O3_O4,eq:O5_O6} imply that, once the contribution of the generalised Hamiltonian in \cref{eq:Heff-NP} to the HQE in \cref{eq:GammaB} has been computed, the corresponding effects induced by the SM QCD-penguin operators can be inferred directly.
Specifically, the matching coefficients up to dimension-six arising from the QCD-penguin operators $Q_{3, \ldots, 6}^{qq_3}$ in \cref{eq:Heff-SM} within the SM can be extracted from the expressions for $P_{i,j}^{(q_1 q_2q_3)} (\rho)$, $G_{i,j}^{(q_1q_2q_3)} (\rho) $, and $D_{i,j}^{(q_1q_2q_3)} (\rho)$, by means of Fierz transformations.
Since we neglect the masses of the light quarks $u,d,s,$ only two mass configurations are relevant, namely case (ii) with $m_{1} = m_{2} = m_c$ and case (iv). 
While the SM QCD-penguin contributions at dimension-three were already known~\cite{Krinner:2013cja}, the subleading power corrections computed here are new and arise as a by-product of our calculation. These results are discussed in \cref{sec:SM_penguin}.
\subsection[Results for the $b \to c \bar u d$ transition]{\texorpdfstring{\boldmath Results for the $b \to c \bar u d$ transition}{Results for the b -> c ubar d transition}}
\label{sec:bcud}
\paragraph{Dimension-three contribution:}
The non-vanishing coefficient functions $P_{i,j}^{(cud)}(\rho)$ in \cref{eq:Fij} exhibit simple symmetry and proportionality relations.
The independent ones read:
\begin{align}
P_{1,1}^{(cud)}(\rho) &= 
1 - 8\rho - 12\rho^2 \log\rho + 8\rho^3 - \rho^4 \,, \label{eq:P11} \\[2mm]
P_{1,3}^{(cud)}(\rho) &=
- 2 \sqrt{\rho}\,\big(
1 + 9\rho + 6 \rho (1 + \rho)\log\rho - 9\rho^2 - \rho^3  
\big) \,. \label{eq:P13}
\end{align}
All other functions follow from the relations
\begin{equation}
\begin{aligned}
& P_{2,2}^{(cud)} =  P_{3,3}^{(cud)} = P_{4,4}^{(cud)} = P_{1,1}^{(cud)} \,,
\qquad
P_{5,5}^{(cud)} = P_{6,6}^{(cud)} = P_{7,7}^{(cud)} = P_{8,8}^{(cud)} = \frac{1}{4}P_{1,1}^{(cud)} \,,
\\[6pt]
&  P_{9,9}^{(cud)} = P_{10,10}^{(cud)} = 12\,P_{1,1}^{(cud)},
\qquad
P_{2,4}^{(cud)} = P_{1,3}^{(cud)} \,,
\qquad
P_{5,7}^{(cud)} = P_{6,8}^{(cud)} = -\frac{1}{2}P_{1,3}^{(cud)} \,,
\\[6pt]
&  
P_{i,j}^{(cud)} = P_{i,j-1}^{(cud)}
\quad \text{for odd } i \text{ and even } j \,,
\end{aligned}
\label{eq:Prellations}
\end{equation}
with all remaining combinations vanishing.
The dimension-three contributions to $b \to c \bar{u} d$ have been recently presented in Ref.~\cite{Meiser:2024zea}.
Taking into account the appropriate change of basis (and with $N_c = 3$) we find full agreement with those results.
\paragraph{Dimension-five contribution:} The contribution of the kinetic operator is obtained from that at dimension-three up to a constant factor, cf.\ \cref{eq:Fij}. 
As for the contribution of the chromomagnetic operator, the independent non-vanishing functions $G_{i,j}^{cud}(\rho)$ in \cref{eq:Fij} read:
\begin{align}
G_{1,1}^{(cud)}(\rho) &= 
-\frac{1}{2}\big(5\rho^4 - 24\rho^3 + 24\rho^2 + 12\rho^2\log\rho - 8\rho + 3\big) \,, \label{eq:G11} \\[2pt]
G_{1,2}^{(cud)}(\rho) &= 
-\frac{1}{2}\big(5\rho^4 - 40\rho^3 + 72\rho^2 + 12\rho^2\log\rho - 56\rho + 19\big) \,, \label{eq:G12} \\[2pt]
G_{1,3}^{(cud)}(\rho) &= 
\frac{1}{3}\sqrt{\rho}\big(13\rho^3 - 27\rho^2 - 6(3\rho^2 - 3\rho + 2)\log\rho + 27\rho - 13\big) \,, \label{eq:G13} \\[2pt]
G_{2,4}^{(cud)}(\rho) &= 
\sqrt{\rho}\big(5\rho^3 - 3\rho^2 - 6(1 + \rho)\rho\log\rho + 3\rho - 5\big) \,, \label{eq:G24} \\[2pt]
G_{4,4}^{(cud)}(\rho) &= 
-\frac{1}{2}\big(5\rho^4 - 8\rho^3 - 24\rho^2 + 12\rho^2\log\rho + 40\rho - 13\big) \,, \label{eq:G44} \\[2pt]
G_{5,5}^{(cud)}(\rho) &= 
-\frac{1}{8}\big(5\rho^4 - 32\rho^3 + 72\rho^2 + 12(\rho-4)\rho\log\rho - 32\rho - 13\big) \,, \label{eq:G55} \\[2pt]
G_{5,7}^{(cud)}(\rho) &= 
-\frac{3}{2}\sqrt{\rho}\big(\rho^3 - 3\rho^2 - 2(\rho^2 - 5\rho - 2)\log\rho - 9\rho + 11\big) \,, \label{eq:G57} \\[2pt]
G_{6,8}^{(cud)}(\rho) &= 
\frac{3}{2}(1-\rho)\sqrt{\rho}\,(\rho^2 - 2\rho\log\rho - 1) \,, \label{eq:G68} \\[2pt]
G_{6,10}^{(cud)}(\rho) &= 
-4\sqrt{\rho}\,(\rho^3 - 6\rho\log\rho + 3\rho - 4) \,, \label{eq:G610} \\[2pt]
G_{8,10}^{(cud)}(\rho) &= 8(1-\rho)^3, \label{eq:G810} \\[2pt]
G_{9,9}^{(cud)}(\rho) &= 
-2\big(15\rho^4 - 64\rho^3 + 24\rho^2 + 12(3\rho+4)\rho\log\rho + 25\big) \,, \label{eq:G99} \\[2pt]
G_{10,10}^{(cud)}(\rho) &= 
-2\big(15\rho^4 - 40\rho^3 - 24\rho^2 + 36\rho^2\log\rho + 72\rho - 23\big) \,. \label{eq:G1010}
\end{align}
All other functions follow from the relations
\begin{equation}
\begin{aligned}
& G_{3,3}^{(cud)} = G_{2,2}^{(cud)} = G_{1,1}^{(cud)},\qquad 
G_{7,7}^{(cud)} = G_{5,5}^{(cud)} \,,\qquad 
G_{8,8}^{(cud)} = G_{6,6}^{(cud)} = \frac{1}{4}G_{1,1}^{(cud)} \,, \\[6pt]
& 
G_{3,2}^{(cud)} = G_{1,4}^{(cud)} = G_{1,3}^{(cud)} \,, \qquad 
G_{3,4}^{(cud)} = G_{4,4}^{(cud)} \,,\qquad 
G_{5,6}^{(cud)} = G_{5,5}^{(cud)} \,, \\[6pt]
& 
 G_{5,8}^{(cud)} = G_{5,7}^{(cud)} \,,\qquad 
G_{5,10}^{(cud)} = 2G_{6,10}^{(cud)} \,,
\qquad
G_{9,6}^{(cud)} = 2G_{6,10}^{(cud)} \,,\\[6pt]
& 
G_{7,8}^{(cud)} = G_{7,7}^{(cud)} \,,\qquad 
G_{7,10}^{(cud)} = 2G_{8,10}^{(cud)} \,,\qquad 
G_{7,6}^{(cud)} = G_{5,7}^{(cud)} \,,
\\[6pt]
& G_{9,8}^{(cud)} = G_{7,10}^{(cud)} \,, \qquad 
G_{9,10}^{(cud)} = G_{9,9}^{(cud)} \, ,
\end{aligned}
\label{eq:Grelations}
\end{equation}
with all remaining combinations vanishing.
\paragraph{Dimension-six contribution:}
For the contribution of the Darwin operator, the independent non-vanishing functions $D_{i,j}^{cud}(\rho)$ in \cref{eq:Fij} read:
\begin{align}
D_{1,1}^{(cud)}(\rho) &= -\frac{2}{3}\big(5\rho^4 - 16\rho^3 + 12\rho^2 + 16\rho - 17 - 12\log\rho\big) \,, \label{eq:D11} \\[2pt]
D_{1,2}^{(cud)}(\rho) &= -\frac{2}{3} \big(5 \rho^4 - 54 \rho^3 + 90 \rho^2 - 50 \rho + 9\big) - 16 (\rho - 1)^3 \log(1 - \rho) \notag \\
& \qquad \qquad + 8 (\rho^3 - 3 \rho^2 + 1) \log\rho \,, \label{eq:D12} 
\end{align}
\begin{align}
D_{1,3}^{(cud)}(\rho) &= \frac{8}{3} \sqrt{\rho} \big(2 \rho^3 - 9 \rho^2 + 18 \rho - 11 - 6 \log\rho\big) \,, \label{eq:D13} \\[2pt]
D_{1,4}^{(cud)}(\rho) &= \frac{16}{3} \sqrt{\rho} \big(\rho^3 - 6 \rho^2 + 12 \rho - 7 - 3 \log\rho\big) \,, \label{eq:D14} \\[2pt]
D_{2,2}^{(cud)}(\rho) &= \frac{2}{3} (1 - \rho) \big(5 \rho^3 - 25 \rho^2 + 11 \rho + 9 + 24 (\rho^2 - 1) \log(1 - \rho) - 12 \rho^2 \log\rho\big) \,, \label{eq:D22} \\[2pt]
D_{2,4}^{(cud)}(\rho) &= \frac{4}{3} \sqrt{\rho} \big(5 \rho^3 - 24 \rho^2 + 27 \rho - 8 + 12 (\rho - 1)^2 \log(1 - \rho) - 6 \rho (\rho - 2) \log\rho\big) \,, \label{eq:D24} \\[2pt]
D_{3,4}^{(cud)}(\rho) &= -\frac{2}{3} \big(5 \rho^4 + 18 \rho^3 - 78 \rho^2 + 118 \rho - 63 + 24 (1 - \rho)^3 \log(1 - \rho) \notag \\ 
& \qquad \qquad + 12 (\rho^3 - 3 \rho^2 - 1) \log\rho\big) \,, \label{eq:D34} \\[2pt]
D_{4,4}^{(cud)}(\rho) &= -\frac{2}{3} \big(5 \rho^4 + 8 \rho^3 - 66 \rho^2 + 112 \rho - 59 + 48 (1 - \rho)^2 \log(1 - \rho) - 24 \rho^2 \log\rho\big) \,, \label{eq:D44} \\[2pt]
D_{5,5}^{(cud)}(\rho) &= \frac{1}{6} \big(-5 \rho^4 + 22 \rho^3 - 54 \rho^2 - 22 \rho + 59 + 12 (5 \rho + 2) \log\rho\big) \,, \label{eq:D55} \\[2pt]
D_{5,6}^{(cud)}(\rho) &= \frac{1}{6} \big(-5 \rho^4 + 26 \rho^3 - 66 \rho^2 - 10 \rho + 55 + 12 (5 \rho + 2) \log\rho\big) \,, \label{eq:D56} \\[2pt]
D_{6,8}^{(cud)}(\rho) &= -\frac{4}{3} \sqrt{\rho} \big(\rho^3 - 6 \rho^2 + 6 \rho - 1 + 6 (\rho - 1)^2 \log(1 - \rho) - 3 \rho (\rho - 2) \log\rho\big) \,, \label{eq:D68} \\[2pt]
D_{6,10}^{(cud)}(\rho) &= -8 \sqrt{\rho} (\rho - 2) (\rho - 1)^2 \,, \label{eq:D610} \\[2pt]
D_{7,8}^{(cud)}(\rho) &= \frac{1}{6} \big(-5 \rho^4 + 18 \rho^3 - 42 \rho^2 - 34 \rho + 63\big) + 2 (5 \rho + 2) \log\rho \,, \label{eq:D78} \\[2pt]
D_{7,10}^{(cud)}(\rho) &= 16 \big(-3 \rho^3 + 7 \rho^2 - 7 \rho + 3 + 2 (\rho - 1)^3 \log(1 - \rho) - \rho^2 (\rho - 3) \log\rho\big) \,, \label{eq:D710} \\[2pt]
D_{8,8}^{(cud)}(\rho) &= \frac{1}{6} (1 - \rho) \big(5 \rho^3 - 21 \rho^2 + 3 \rho + 13 + 24 (\rho^2 - 1) \log(1 - \rho) - 12 \rho^2 \log\rho\big) \,, \label{eq:D88} \\[2pt]
D_{8,10}^{(cud)}(\rho) &= \frac{4}{3} \big(-17 \rho^3 + 45 \rho^2 - 51 \rho + 23 + 12 (\rho - 1)^3 \log(1 - \rho) - 6 \rho^2 (\rho - 3) \log\rho\big) \,, \label{eq:D810} \\[2pt]
D_{9,9}^{(cud)}(\rho) &= -8 \big(5 \rho^4 - 14 \rho^3 - 2 \rho^2 + 14 \rho - 3 + 4 (5 \rho - 2) \log\rho\big) \,, \label{eq:D99} \\[2pt]
D_{9,10}^{(cud)}(\rho) &= -\frac{8}{3} \big(15 \rho^4 - 38 \rho^3 - 66 \rho^2 + 150 \rho - 61 + 12 (5 \rho - 2) \log\rho\big) \,, \label{eq:D910} \\[2pt]
D_{10,10}^{(cud)}(\rho) &= -\frac{8}{3} \big(15 \rho^4 - 10 \rho^3 - 108 \rho^2 + 234 \rho - 131 + 24 (\rho - 1)^2 (\rho + 5) \log(1 - \rho) \notag \\ 
& \qquad \qquad - 12 \rho^2 (\rho + 3) \log\rho\big) \,. \label{eq:D1010}
\end{align}
All other functions follow from the relations
\begin{equation}
\begin{aligned}
& D_{3,3}^{(cud)} = D_{1,1}^{(cud)} \,, \qquad 
D_{7,7}^{(cud)} = D_{5,5}^{(cud)} \,, \qquad 
D_{6,6}^{(cud)} = \frac{1}{4}D_{2,2}^{(cud)} \,,\\[6pt]
&  
D_{5,7}^{(cud)} = -\frac{1}{4}D_{1,3}^{(cud)} \,, \qquad 
D_{5,8}^{(cud)} = -\frac{1}{4}D_{1,3}^{(cud)} \,, \qquad D_{5,10}^{(cud)} = \frac{2(\rho-1)}{\rho-2}D_{6,10}^{(cud)} \,,\\[6pt]
&  
D_{i,j}^{(cud)} = D_{i-1,j+1}^{(cud)} \qquad \text{for even } i \text{ and odd } j \,,
\end{aligned}
\label{eq:Drelations}
\end{equation}
with all remaining combinations vanishing.
\subsection[Results for the $b \to c \bar c s$ transition]{\texorpdfstring{\boldmath Results for the $b \to c \bar c s$ transition}{Results for the b -> c cbar s transition}}
\label{sec:bccs}
\paragraph{Dimension-three contribution:} 
Introducing the shorthand notation $ \Delta_{\rho} \equiv \sqrt{1 - 4\rho},$ and $ 
\Lambda_{\rho} \equiv \log\left[\frac{1 + \Delta_{\rho}}{1 - \Delta_{\rho}}\right],
$ the independent functions $P_{i,j}^{(ccs)}(\rho)$ in \cref{eq:Fij} read:
\begin{align}
P_{1,1}^{(ccs)}(\rho) &= \Delta_{\rho}\left(1 - 14\rho - 2\rho^2 - 12\rho^3 \right) + 24\rho^2 (1 - \rho^2) \Lambda_{\rho} \,, \label{eq:P11ccs} \\[6pt]
P_{1,3}^{(ccs)}(\rho) &= 2\sqrt{\rho} \left[ \Delta_{\rho}\left(-1 - 5\rho + 6\rho^2 \right) + 6\rho (1 - 2\rho + 2\rho^2) \Lambda_{\rho} \right] \,. \label{eq:P13ccs}
\end{align}
All other functions follow from the relations
\begin{equation}
\begin{aligned}
& P_{2,2}^{(ccs)} = P_{3,3}^{(ccs)} = P_{4,4}^{(ccs)} = P_{1,1}^{(ccs)} \,,
\qquad
P_{5,5}^{(ccs)} = P_{6,6}^{(ccs)} = P_{7,7}^{(ccs)} = P_{8,8}^{(ccs)} = \frac{1}{4}P_{1,1}^{(ccs)} \,,
\\[6pt]
& P_{9,9}^{(ccs)} = P_{10,10}^{(ccs)} = 12\,P_{1,1}^{(ccs)} \,,
\qquad
P_{2,4}^{(ccs)} = P_{1,3}^{(ccs)} \,,
\qquad
P_{5,7}^{(ccs)} = P_{6,8}^{(ccs)} = -\frac{1}{2}P_{1,3}^{(ccs)} \,,
\\[6pt]
& P_{1,5}^{(ccs)} = P_{3,7}^{(ccs)} = P_{2,6}^{(ccs)} = P_{4,8}^{(ccs)} = \sqrt{\rho} \, P_{1,3}^{(ccs)} \,,
\qquad
P_{1,7}^{(ccs)} = P_{3,5}^{(ccs)} = P_{2,8}^{(ccs)} = P_{4,6}^{(ccs)} = \frac{1}{2} P_{1,3}^{(ccs)} \,,
\\[6pt]
& P_{1,9}^{(ccs)} = P_{2,10}^{(ccs)} = -6 P_{1,3}^{(ccs)} \,,
\qquad
P_{3,9}^{(ccs)} = P_{4,10}^{(ccs)} = 12\sqrt{\rho} P_{1,3}^{(ccs)} \,,
\\[6pt]
& P_{i,j}^{(ccs)} = P_{i,j-1}^{(ccs)}
\qquad \text{for odd } i \text{ and even } j \,,
\end{aligned}
\label{eq:Prelationsccs}
\end{equation}
with all remaining combinations vanishing.

\paragraph{Dimension-five contribution:}
The contribution of the kinetic operator is obtained from that at dimension-three up to a constant factor, cf.\ \cref{eq:Fij}. 
As for the contribution of the chromomagnetic operator, the independent non-vanishing functions $G_{i,j}^{ccs}(\rho)$ in \cref{eq:Fij} read:
\begin{align}
G_{1,1}^{(ccs)}(\rho) &= -\frac{1}{2\Delta_{\rho}} \Big[ 3 - 22\rho + 50\rho^2 + 20\rho^3 - 240\rho^4 + 24\Delta_{\rho}\rho^2(-1 + 5\rho^2)\Lambda_{\rho} \Big] \,, \label{eq:G11ccs} \\[2pt]
G_{1,2}^{(ccs)}(\rho) &= \frac{1}{2\Delta_{\rho}} \Big[ -19 + 78\rho - 66\rho^2 + 172\rho^3 + 240\rho^4 - 24\Delta_{\rho}\rho(-2 - \rho + 4\rho^2 + 5\rho^3)\Lambda_{\rho} \Big] \,, \label{eq:G12ccs} \\[2pt]
G_{1,3}^{(ccs)}(\rho) &= \frac{\sqrt{\rho}}{3\Delta_{\rho}} \Big[ -13 + 23\rho + 266\rho^2 - 600\rho^3 + 6\Delta_{\rho}(2 - 3\rho - 18\rho^2 + 50\rho^3)\Lambda_{\rho} \Big] \,, \label{eq:G13ccs} \\[2pt]
G_{1,4}^{(ccs)}(\rho) &= \frac{\sqrt{\rho}}{3\Delta_{\rho}} \Big[ -13 + 41\rho + 302\rho^2 - 1032\rho^3 + 6\Delta_{\rho}(2 - 9\rho - 18\rho^2 + 86\rho^3)\Lambda_{\rho} \Big] \,, \label{eq:G14ccs} \\[2pt]
G_{1,5}^{(ccs)}(\rho) &= \frac{\rho}{\Delta_{\rho}} \Big[ 10 - 71\rho + 154\rho^2 - 120\rho^3 + 6\Delta_{\rho}(-1 + 5\rho - 12\rho^2 + 10\rho^3)\Lambda_{\rho} \Big] \,, \label{eq:G15ccs} \\[2pt]
G_{1,6}^{(ccs)}(\rho) &= \frac{\rho}{\Delta_{\rho}} \Big[ 7 - 53\rho + 130\rho^2 - 120\rho^3 + 6\Delta_{\rho}\rho(1 - 10\rho + 10\rho^2)\Lambda_{\rho} \Big] \,, \label{eq:G16ccs} \\[2pt]
G_{1,7}^{(ccs)}(\rho) &= \frac{\sqrt{\rho}}{2\Delta_{\rho}} \Big[ -5 + 13\rho + 22\rho^2 + 24\rho^3 - 6\Delta_{\rho}\rho(-3 + 2\rho + 2\rho^2)\Lambda_{\rho} \Big] \,, \label{eq:G17ccs} \\[2pt]
G_{1,8}^{(ccs)}(\rho) &= -\frac{3\sqrt{\rho}}{2\Delta_{\rho}} \Big[ 9 - 37\rho + 10\rho^2 - 24\rho^3 + 2\Delta_{\rho}(-2 - \rho - 2\rho^2 + 6\rho^3)\Lambda_{\rho} \Big] \,, \label{eq:G18ccs}
\end{align}
\begin{align}
G_{1,9}^{(ccs)}(\rho) &= \frac{2\sqrt{\rho}}{\Delta_{\rho}} \Big[ -17 + 73\rho - 98\rho^2 + 312\rho^3 - 6\Delta_{\rho}\rho(-7 - 6\rho + 26\rho^2)\Lambda_{\rho} \Big] \,, \label{eq:G19ccs} \\[2pt]
G_{1,10}^{(ccs)}(\rho) &= -\frac{6\sqrt{\rho}}{\Delta_{\rho}} \Big[ -27 + 119\rho - 14\rho^2 - 120\rho^3 + 2\Delta_{\rho}(6 - 5\rho + 6\rho^2 + 30\rho^3)\Lambda_{\rho} \Big] \,, \label{eq:G110ccs}
\\[2pt]
G_{2,4}^{(ccs)}(\rho) &= \frac{\sqrt{\rho}}{\Delta_{\rho}} \Big[ -5 + 19\rho + 34\rho^2 - 120\rho^3 + 6\Delta_{\rho}\rho(1 - 2\rho + 10\rho^2)\Lambda_{\rho} \Big] \,, \label{eq:G24ccs} \\[2pt]
G_{2,8}^{(ccs)}(\rho) &= \frac{\sqrt{\rho}}{2\Delta_{\rho}} \Big[ 11 - 49\rho + 26\rho^2 - 24\rho^3 + 6\Delta_{\rho}\rho(-3 - 2\rho + 2\rho^2)\Lambda_{\rho} \Big] \,, \label{eq:G28ccs} \\[2pt]
G_{2,10}^{(ccs)}(\rho) &= \frac{2\sqrt{\rho}}{\Delta_{\rho}} \Big[ -1 + 11\rho - 94\rho^2 + 264\rho^3 + 6\Delta_{\rho}\rho(1 + 6\rho - 22\rho^2)\Lambda_{\rho} \Big] \,, \label{eq:G210ccs} \\[2pt]
G_{3,4}^{(ccs)}(\rho) &= \frac{1}{2\Delta_{\rho}} \Big[ 13 - 106\rho + 206\rho^2 - 20\rho^3 + 240\rho^4 - 24\Delta_{\rho}\rho^2(-1 + 5\rho^2)\Lambda_{\rho} \Big] \,, \label{eq:G34ccs} \\[2pt]
G_{3,5}^{(ccs)}(\rho) &= \frac{\sqrt{\rho}}{2\Delta_{\rho}} \Big[ -5 + 13\rho + 22\rho^2 + 24\rho^3 - 6\Delta_{\rho}\rho(-3 + 2\rho + 2\rho^2)\Lambda_{\rho} \Big] \,, \label{eq:G35ccs} \\[2pt]
G_{3,6}^{(ccs)}(\rho) &= \frac{\sqrt{\rho}}{6\Delta_{\rho}} \Big[ -17 + 73\rho - 98\rho^2 + 312\rho^3 - 6\Delta_{\rho}(2 - 15\rho - 6\rho^2 + 26\rho^3)\Lambda_{\rho} \Big] \,, \label{eq:G36ccs} \\[2pt]
G_{3,7}^{(ccs)}(\rho) &= \frac{\rho}{\Delta_{\rho}} \Big[ 10 - 71\rho + 154\rho^2 - 120\rho^3 + 6\Delta_{\rho}(-1 + 5\rho - 12\rho^2 + 10\rho^3)\Lambda_{\rho} \Big] \,, \label{eq:G37ccs} \\[2pt]
G_{3,8}^{(ccs)}(\rho) &= \frac{\rho}{\Delta_{\rho}} \Big[ 13 - 89\rho + 178\rho^2 - 120\rho^3 + 6\Delta_{\rho}(-2 + 9\rho - 14\rho^2 + 10\rho^3)\Lambda_{\rho} \Big] \,, \label{eq:G38ccs} \\[2pt]
G_{3,9}^{(ccs)}(\rho) &= \frac{12\rho}{\Delta_{\rho}} \Big[ -2 + \rho + 58\rho^2 - 120\rho^3 + 2\Delta_{\rho}(1 - \rho - 12\rho^2 + 30\rho^3)\Lambda_{\rho} \Big] \,, \label{eq:G39ccs} \\[2pt]
G_{3,10}^{(ccs)}(\rho) &= \frac{12\rho}{\Delta_{\rho}} \Big[ -11 + 55\rho - 14\rho^2 - 120\rho^3 + 2\Delta_{\rho}(2 - 5\rho + 6\rho^2 + 30\rho^3)\Lambda_{\rho} \Big] \,, \label{eq:G310ccs} \\[2pt]
G_{5,5}^{(ccs)}(\rho) &= \frac{1}{8\Delta_{\rho}} \Big[ 13 - 34\rho - 34\rho^2 - 212\rho^3 + 240\rho^4 - 24\Delta_{\rho}\rho(2 - \rho - 4\rho^2 + 5\rho^3)\Lambda_{\rho} \Big] \,, \label{eq:G55ccs} \\[2pt]
G_{5,7}^{(ccs)}(\rho) &= -\frac{3\sqrt{\rho}}{2\Delta_{\rho}} \Big[ 11 - 49\rho + 26\rho^2 - 24\rho^3 + 2\Delta_{\rho}(-2 - \rho - 6\rho^2 + 6\rho^3)\Lambda_{\rho} \Big] \,, \label{eq:G57ccs} \\[2pt]
G_{5,10}^{(ccs)}(\rho) &= 4\sqrt{\rho} \big[ \Delta_{\rho}(8 + \rho + 6\rho^2) + 6\rho(-3 + 2\rho^2)\Lambda_{\rho} \big] \,, \label{eq:G510ccs} \\[2pt]
G_{6,8}^{(ccs)}(\rho) &= \frac{3\sqrt{\rho}}{2\Delta_{\rho}} \Big[ -1 + 5\rho - 10\rho^2 + 24\rho^3 + 2\Delta_{\rho}\rho(1 + 2\rho - 6\rho^2)\Lambda_{\rho} \Big] \,, \label{eq:G68ccs} \\[2pt]
G_{6,10}^{(ccs)}(\rho) &= -4\sqrt{\rho} \big[ \Delta_{\rho}(-4 + \rho + 6\rho^2) + 6(\rho + 2\rho^3)\Lambda_{\rho} \big] \,, \label{eq:G610ccs} \\[2pt]
G_{7,10}^{(ccs)}(\rho) &= 4 \big[ \Delta_{\rho}(4 - 7\rho + 6\rho^2) + 6\rho(-1 + 2\rho^2)\Lambda_{\rho} \big] \,, \label{eq:G710ccs} \\[2pt]
G_{8,10}^{(ccs)}(\rho) &= 8\Delta_{\rho}^{3} \,, \label{eq:G810ccs} \\[2pt]
G_{9,9}^{(ccs)}(\rho) &= \frac{2}{\Delta_{\rho}} \Big[ -25 + 122\rho - 166\rho^2 + 132\rho^3 + 720\rho^4 - 24\Delta_{\rho}\rho(-2 - 3\rho + 4\rho^2 + 15\rho^3)\Lambda_{\rho} \Big] \,, \label{eq:G99ccs} \\[2pt]
G_{9,10}^{(ccs)}(\rho) &= \frac{2}{\Delta_{\rho}} \Big[ -25 - 22\rho + 314\rho^2 + 516\rho^3 + 720\rho^4 - 72\Delta_{\rho}\rho(-2 - \rho + 4\rho^2 + 5\rho^3)\Lambda_{\rho} \Big] \,, \label{eq:G910ccs} \\[2pt]
G_{10,10}^{(ccs)}(\rho) &= \frac{2}{\Delta_{\rho}} \Big[ 23 - 190\rho + 362\rho^2 - 60\rho^3 + 720\rho^4 - 72\Delta_{\rho}\rho^2(-1 + 5\rho^2)\Lambda_{\rho} \Big] \,. \label{eq:G1010ccs}
\end{align}
All other functions follow from the relations
\begin{equation}
\begin{aligned}
& G_{2,2}^{(ccs)} = G_{1,1}^{(ccs)} \,,\qquad G_{2,6}^{(ccs)} = G_{1,6}^{(ccs)} \,,\qquad G_{3,3}^{(ccs)} = G_{1,1}^{(ccs)} \,,\qquad G_{4,4}^{(ccs)} = G_{3,4}^{(ccs)} \,,\\[6pt]
& G_{3,5}^{(ccs)} = G_{1,7}^{(ccs)} \,,\qquad G_{4,6}^{(ccs)} = \frac{1}{2}G_{2,4}^{(ccs)} \,, 
\qquad G_{4,8}^{(ccs)} = \sqrt{\rho} \, G_{2,4}^{(ccs)} \,,\qquad G_{4,10}^{(ccs)} = 12\sqrt{\rho}\, G_{2,4}^{(ccs)} \,, \\[6pt]
&  G_{5,6}^{(ccs)} = G_{5,5}^{(ccs)} \,,\qquad G_{5,8}^{(ccs)} = G_{5,7}^{(ccs)} \,,\qquad G_{6,6}^{(ccs)} = \frac{1}{4}G_{3,3}^{(ccs)} \,, \qquad 
G_{7,8}^{(ccs)} = G_{7,7}^{(ccs)} = G_{5,5}^{(ccs)} \,, \\[6pt]
&  G_{8,8}^{(ccs)} = G_{6,6}^{(ccs)} \,,\qquad G_{i,j}^{(ccs)} = G_{i-1,j+1}^{(ccs)} \qquad \text{for even } i \text{ and odd } j \,,
\end{aligned}
\label{eq:Grelationsccs}
\end{equation}
with all remaining combinations vanishing.
\paragraph{Dimension-six contribution:}
For the contribution of the Darwin operator, the independent non-vanishing functions $D_{i,j}^{ccs}(\rho)$ in \cref{eq:Fij} read:
\begin{align}
D_{1,1}^{(ccs)}(\rho) &= -\frac{2}{3} \Big[ \Delta_{\rho} (-17 - 8\rho + 22\rho^2 + 60\rho^3) + 12 (1 - \rho - 2\rho^2 + 2\rho^3 + 10\rho^4) \Lambda_{\rho} \Big] \,, \label{eq:D11ccs} \\[2pt]
D_{1,2}^{(ccs)}(\rho)& =
- \frac{2}{3} \Delta_\rho \left[33 - 46 \rho + 106 \rho^2 + 60 \rho^3 + 24 \log (1- 4 \rho) - 24 \log \rho \right]
\nonumber \\
&  \qquad \qquad + 8 \left[3 - 2 \rho + 4 \rho^2 - 16 \rho^3 - 10\rho^4 \right] \Lambda_\rho 
\label{eq:D12ccs} \\[2pt]
D_{1,3}^{(ccs)}(\rho) &= \frac{8}{3} \sqrt{\rho} \Big[ \Delta_{\rho} (-11 - 7\rho + 30\rho^2) + 6 (1 - 4\rho^2 + 10\rho^3) \Lambda_{\rho} \Big] \,, \label{eq:D13ccs} \\[2pt]
D_{1,4}^{(ccs)}(\rho) & = \frac{4}{3} \sqrt{\rho} \Big[\Delta_\rho (-22 - 23 \rho + 114 \rho^2 + 12 \log(1 - 4 \rho) - 
      12 \log(\rho))  \nonumber \\
& \qquad \qquad + 6 (-1 + 6 \rho - 14 \rho^2 + 38 \rho^3) \Lambda_\rho \Big]  \,,
\label{eq:D14ccs} \\[2pt]
D_{1,5}^{(ccs)}(\rho) &= \frac{4}{3} \rho \Big[ \Delta_{\rho} (7 - 31\rho + 30\rho^2) + 3 (-1 + 8\rho - 24\rho^2 + 20\rho^3) \Lambda_{\rho} \Big] \,, \label{eq:D15ccs} \\[2pt]
D_{1,7}^{(ccs)}(\rho) &= -\frac{4}{3} \sqrt{\rho} \Big[ 2\Delta_{\rho} (2 - 5\rho + 6\rho^2) + 3 (-1 + 4\rho - 8\rho^2 + 8\rho^3) \Lambda_{\rho} \Big] \,, \label{eq:D17ccs} \\[2pt]
D_{1,8}^{(ccs)} (\rho)  & = \frac{2}{3} \sqrt{\rho} \Big[ \Delta_\rho (7 + 29 \rho - 42 \rho^2 + 12 \log(1 - 4 \rho) - 12 \log \rho)
\notag \\[2pt]
& \qquad \qquad  - 12 (1 + \rho - 6 \rho^2 + 7 \rho^3) \Lambda_\rho \Big] \,, \label{eq:D18ccs} \\[2pt]
D_{1,9}^{(ccs)}(\rho) &= -16 \sqrt{\rho} \Big[ 2\Delta_{\rho} (-1 + \rho + 6\rho^2) + 3 (1 - 4\rho + 8\rho^3) \Lambda_{\rho} \Big] \,, \label{eq:D19ccs} \\[2pt]
D_{1,10}^{(ccs)} (\rho)  & = 8 \sqrt{\rho} \Big[-\Delta_\rho (3 + 17 \rho + 30 \rho^2 + 12 \log(1 - 4 \rho) - 
      12 \log \rho)  \nonumber \\ 
& \qquad \qquad + 12 (1 + \rho - 2 \rho^2 - 5 \rho^3) \Lambda_\rho \Big] \,, \label{eq:D110ccs} \\[2pt]
D_{2,2}^{(ccs)}(\rho) &= \frac{2}{3} \Delta_{\rho} \big[9 + 22\rho - 34\rho^2 - 60\rho^3 - 24\log(1 - 4\rho) + 24\log\rho\big] \nonumber \\
& \qquad \qquad - 16 \big[-1 + 2\rho + \rho^2 + 2\rho^3 + 5\rho^4\big] \Lambda_{\rho} \,, \label{eq:D22ccs} \\[2pt]
D_{2,4}^{(ccs)}(\rho) &= \frac{4}{3} \sqrt{\rho} \Big[ \Delta_{\rho} \big(-8 + 11\rho + 30\rho^2 + 12\log(1 - 4\rho) - 12\log\rho\big) \nonumber \\
& \qquad \qquad + 12 \big(-1 + \rho + \rho^2 + 5\rho^3\big) \Lambda_{\rho} \Big] \,, \label{eq:D24ccs}
\end{align}
\begin{align}
D_{2,6}^{(ccs)}(\rho) &= \frac{4}{3} \rho \Big[ \Delta_{\rho} \big(7 - 7\rho + 30\rho^2 + 12\log(1 - 4\rho) - 12\log\rho\big) \notag \\
& \qquad \qquad + 12 \big(-1 + \rho - 2\rho^2 + 5\rho^3\big) \Lambda_{\rho} \Big] \,, \label{eq:D26ccs} \\[2pt]
D_{2,8}^{(ccs)}(\rho) &= \frac{2}{3} \sqrt{\rho} \Big[ \Delta_{\rho} \big(4 + 5\rho - 6\rho^2 + 12\log(1 - 4\rho) - 12\log\rho\big) - 12 (-1 + \rho)^2 (1 + \rho) \Lambda_{\rho} \Big] \,, \label{eq:D28ccs} \\[2pt]
D_{2,10}^{(ccs)}(\rho) &= 8 \sqrt{\rho} \Big[ \Delta_{\rho} \big(4 - 9\rho - 18\rho^2 - 12\log(1 - 4\rho) + 12\log\rho\big) \nonumber \\
& \qquad \qquad - 12 \big(-1 + \rho + \rho^2 + 3\rho^3\big) \Lambda_{\rho} \Big] \,, \label{eq:D210ccs} \\[2pt]
D_{3,4}^{(ccs)}(\rho) &= -\frac{2}{3} \Delta_{\rho} \big[-59 + 106\rho + 10\rho^2 + 60\rho^3 + 48(1-2\rho)\log(1 - 4\rho) + 48(-1 + 2\rho)\log\rho\big] \nonumber \\
& \qquad \qquad + 16 \big(2 - 8\rho + 2\rho^2 - 5\rho^4\big) \Lambda_{\rho} \,, \label{eq:D34ccs} \\[2pt]
D_{3,6}^{(ccs)}(\rho) &= 2 \sqrt{\rho} \Big[ \Delta_{\rho} \big(2 + 15\rho - 18\rho^2 + 4\log(1 - 4\rho) - 4\log\rho\big) \notag \\
& \qquad \qquad - 2 \big(3 + 2\rho - 18\rho^2 + 18\rho^3\big) \Lambda_{\rho} \Big] \,, \label{eq:D36ccs} \\[2pt]
D_{3,8}^{(ccs)}(\rho) &= \frac{2}{3} \rho \Big[ \Delta_{\rho} \big(11 - 32\rho + 60\rho^2 + 24\log(1 - 4\rho) - 24\log\rho\big) \nonumber \\
& \qquad \qquad + 6 \big(-7 + 16\rho - 14\rho^2 + 20\rho^3\big) \Lambda_{\rho} \Big] \,, \label{eq:D38ccs} \\[2pt]
D_{3,9}^{(ccs)}(\rho) &= 16 \rho \Big[ \Delta_{\rho} (-2 - 13\rho + 30\rho^2) + (3 - 36\rho^2 + 60\rho^3) \Lambda_{\rho} \Big] \,, \label{eq:D39ccs} \\[2pt]
D_{3,10}^{(ccs)}(\rho) &= 8 \rho \Big[ \Delta_{\rho} \big(-25 + 40\rho + 60\rho^2 + 24\log(1 - 4\rho) - 24\log\rho\big) \nonumber \\
& \qquad \qquad + 6 \big(-3 + 10\rho^2 + 20\rho^3\big) \Lambda_{\rho} \Big] \,, \label{eq:D310ccs} \\[2pt]
D_{5,5}^{(ccs)}(\rho) &= \frac{1}{6} \Big[ -\Delta_{\rho} (-59 + 16\rho - 50\rho^2 + 60\rho^3) - 12 (2 + \rho + 4\rho^2 - 10\rho^3 + 10\rho^4) \Lambda_{\rho} \Big] \,, \label{eq:D55ccs}\\[2pt]
D_{5,6}^{(ccs)}(\rho) &= \frac{1}{6} \Delta_{\rho} \big[51 - 2\rho + 38\rho^2 - 60\rho^3 - 24\log(1 - 4\rho) + 24\log\rho\big] \nonumber \\
& \qquad \qquad - 2 \big(-1 + 6\rho + 8\rho^2 - 8\rho^3 + 10\rho^4\big) \Lambda_{\rho} \,, \label{eq:D56ccs} \\[2pt]
D_{5,7}^{(ccs)}(\rho) &= -\frac{2}{3} \sqrt{\rho} \Big[ \Delta_{\rho} (-11 - 4\rho + 12\rho^2) + 6 (1 - 2\rho^2 + 4\rho^3) \Lambda_{\rho} \Big] \,, \label{eq:D57ccs} \\[2pt]
D_{5,8}^{(ccs)}(\rho) &= \frac{8}{3} \sqrt{\rho} \Big[ \Delta_{\rho} \big(2 + \rho - 3\rho^2 - 3\log(1 - 4\rho) + 3\log\rho\big) + 3 \big(1 - 3\rho + \rho^2 - 2\rho^3\big) \Lambda_{\rho} \Big] \,, \label{eq:D58ccs} \\[2pt]
D_{5,10}^{(ccs)}(\rho) &= 8 \sqrt{\rho} \Big[ \Delta_{\rho} (3 - 5\rho + 6\rho^2) + 12(-1 + \rho)\rho^2 \Lambda_{\rho} \Big] \,, \label{eq:D510ccs} \\[2pt]
D_{6,8}^{(ccs)}(\rho) &= \frac{4}{3} \sqrt{\rho} \Big[ \Delta_{\rho} \big(1 - 4\rho - 6\rho^2 - 6\log(1 - 4\rho) + 6\log\rho\big) - 6 \big(-1 + \rho + \rho^2 + 2\rho^3\big) \Lambda_{\rho} \Big] \,, \label{eq:D68ccs} \\[2pt]
D_{6,10}^{(ccs)}(\rho) &= -8 \sqrt{\rho} \Big[ \Delta_{\rho} (-2 + \rho + 6\rho^2) + 12\rho^3 \Lambda_{\rho} \Big] \,, \label{eq:D610ccs} \\[2pt]
D_{7,8}^{(ccs)}(\rho) &= \frac{1}{6} \Delta_{\rho} \big[55 - 18\rho + 38\rho^2 - 60\rho^3 - 24\log(1 - 4\rho) + 24\log\rho\big] \nonumber \\
& \qquad \qquad - 2 \big(-1 + 6\rho + 8\rho^2 - 8\rho^3 + 10\rho^4\big) \Lambda_{\rho} \,, \label{eq:D78ccs} \\[2pt]
D_{7,10}^{(ccs)}(\rho) &= \frac{8}{3} \Big[ 2\Delta_{\rho} \big(11 - 17\rho + 12\rho^2 + 3(-1 + 4\rho)\log(1 - 4\rho) + 3(1-4\rho)\log\rho\big) \nonumber \\
& \qquad \qquad + (3 - 42\rho + 48\rho^3) \Lambda_{\rho} \Big] \,, \label{eq:D710ccs}
\end{align}
\begin{align}
D_{8,8}^{(ccs)}(\rho) &= \frac{1}{6} \Delta_{\rho} \big[13 + 6\rho - 34\rho^2 - 60\rho^3 - 24\log(1 - 4\rho) + 24\log\rho\big] \nonumber \\
& \qquad \qquad - 4 \big[-1 + 2\rho + \rho^2 + 2\rho^3 + 5\rho^4\big] \Lambda_{\rho} \,, \label{eq:D88ccs} \\[2pt]
D_{8,10}^{(ccs)}(\rho) &= \frac{4}{3} \Delta_{\rho} \big[23 - 56\rho + 12\rho^2 + 12(-1 + 4\rho)\log(1 - 4\rho) + 12(1-4\rho)\log\rho\big] \nonumber \\
& \qquad \qquad + 16 \big(1 - 6\rho + 3\rho^2 + 2\rho^3\big) \Lambda_{\rho} \,, \label{eq:D810ccs} \\[2pt]
D_{9,9}^{(ccs)}(\rho) &= -8 \Big[ \Delta_{\rho} (-3 - 16\rho + 46\rho^2 + 60\rho^3) + 4 (2 - 5\rho - 12\rho^2 + 18\rho^3 + 30\rho^4) \Lambda_{\rho} \Big] \,, \label{eq:D99ccs} \\[2pt]
D_{9,10}^{(ccs)}(\rho) &= -\frac{8}{3} \Delta_{\rho} \big[-5 + 134\rho + 270\rho^2 + 180\rho^3 - 24(-5 + 8\rho)\log(1 - 4\rho) \nonumber \\
& \qquad \qquad + 24(-5 + 8\rho)\log\rho\big] - 32 \big(-13 + 30\rho - 24\rho^2 + 40\rho^3 + 30\rho^4\big) \Lambda_{\rho} \,, \label{eq:D910ccs} \\[2pt]
D_{10,10}^{(ccs)}(\rho) &= -\frac{8}{3} \Delta_{\rho} \big[-131 + 206\rho + 54\rho^2 + 180\rho^3 - 24(-5 + 8\rho)\log(1 - 4\rho) \nonumber \\
& \qquad \qquad + 24(-5 + 8\rho)\log\rho\big] - 64 \big[-5 + 18\rho - 3\rho^2 + 2\rho^3 + 15\rho^4\big] \Lambda_{\rho} \,. \label{eq:D1010ccs}
\end{align}
All other functions follow from the relations
\begin{equation}
\begin{aligned}
& D_{3,3}^{(ccs)} = D_{1,1}^{(ccs)} \,,\qquad 
D_{4,4}^{(ccs)} = D_{3,4}^{(ccs)} \,, \qquad
D_{3,5}^{(ccs)} = D_{1,7}^{(ccs)} \,,\qquad D_{3,7}^{(ccs)} = D_{1,5}^{(ccs)} \,, \\[6pt]
& D_{6,6}^{(ccs)} = \frac{1}{4}D_{2,2}^{(ccs)} \,, \qquad
D_{4,6}^{(ccs)} = \frac{1}{2}D_{2,4}^{(ccs)} \,,\qquad D_{4,8}^{(ccs)} = \sqrt{\rho}\,D_{2,4}^{(ccs)} \,,\\[6pt]
&  D_{4,10}^{(ccs)} = 12 \sqrt{\rho} \, D_{2,4}^{(ccs)} \,,\qquad D_{2,6}^{(ccs)} = D_{1,6}^{(ccs)} \,, \qquad D_{7,7}^{(ccs)} = D_{5,5}^{(ccs)}\,,\\[6pt]
& D_{i,j}^{(ccs)} = D_{i-1,j+1}^{(ccs)} \qquad \text{for even } i \text{ and odd } j \,,
\end{aligned}
\label{eq:Drelationsccs}
\end{equation}
with all remaining combinations vanishing.
\section{Summary and Outlook}
\label{Sec:outlook}
In this work, we have computed, within the framework of the HQE, the contribution to the total $B$-meson decay width arising from the most general effective Hamiltonian describing possible BSM effects in non-leptonic $b$-quark decays $b \to q_1 \bar q_2 q_3$, with $q_1, q_2 = u,c$ and $q_3 = d,s$. 
Specifically, we have derived analytic expressions at LO in QCD for all matching coefficients of two-quark operators in the HQE, up to mass-dimension-six, including the leading-power results at dimension-three and the coefficients of the chromomagnetic and Darwin operators at dimension-five and six, respectively. \\
At dimension-six, the Darwin operator mixes with four-quark operators under renormalisation, and a consistent treatment of this mixing is required to subtract the IR divergences that appear in the Darwin coefficients from the emission of a soft gluon from one of the light-quark propagators $q = u,d,s$, see Refs.~\cite{Lenz:2020oce, Mannel:2020fts}. To regularise these divergences, we have considered two schemes: introducing a mass regulator for the light quarks in four dimensions, and employing dimensional regularisation with $D = 4 - 2\epsilon$ for massless light quarks. The full agreement between these two approaches provided a non-trivial and robust cross-check of our results. \\
The subtraction of all IR-divergent terms additionally required the computation of weak-annihilation contributions, which enter the matching of four-quark operators at dimension-six and were previously missing. Our results thus
complete the calculation of BSM effects in non-leptonic, tree-level, $b$-quark decays relevant for $B$ meson lifetimes and lifetime ratios such as $\tau(B^0_s)/\tau(B^0_d)$. \\
As a by-product of our calculation, we have also determined the SM matching coefficients up to dimension-six originating from the QCD-penguin operators, including both the interference between current-current and penguin operators and the contributions quadratic in the penguin operators. Owing to the suppression of the QCD-penguin Wilson coefficients within the SM, these effects are typically regarded as corrections of order $\alpha_s$ and $\alpha_s^2$ in the strong coupling, respectively. 
Our results reproduce the known expressions at dimension-three and provide new results for the coefficients of the chromomagnetic operator at dimension-five and the Darwin operator at dimension-six, thereby improving the theoretical precision of SM predictions for $B$-meson lifetimes and lifetimes ratios.\\
Since our analysis included all relevant non-leptonic decay channels, namely the CKM-favoured transitions $b\to c\bar{u}d$ and $b\to c\bar{c}s$, as well as the CKM-suppressed modes, e.g.\ $b\to u\bar{c}s$ and $b\to u\bar{u}d$, this work provides a comprehensive description of non-leptonic $b$-quark decays within the HQE framework in the presence of generic BSM operators. Our results pave the way for global phenomenological analyses that combine lifetime observables with other precision flavour data, such as mixing parameters and CP asymmetries, to improve constraints on BSM Wilson coefficients. First steps in this direction have already been taken in e.g.\ Refs.~\cite{Meiser:2024zea,Lenz:2022pgw}.
Such studies may also help clarify the origin of the tensions observed in several colour-allowed non-leptonic $B$-meson decays when comparing experimental measurements with theoretical predictions based on QCD factorisation.\\
These results can also be mapped onto specific BSM scenarios, such as $W'$ models, diquarks, or extended Higgs sectors, allowing for concrete interpretations of the observed tensions and consistency checks with collider bounds.
Beyond these direct applications, the generic operator basis considered in this work can appear in alternative scenarios with modified colour structures.
After accounting for these colour factors, our expressions can be used in such scenarios as well, see e.g.\ Ref.~\cite{Mohamed:2025zgx}.\\
Further theoretical improvements, including the calculation of NLO-QCD corrections to the Darwin and chromomagnetic contributions, as well as refined lattice QCD determinations of the relevant non-perturbative matrix elements, 
see e.g.\ Refs.~\cite{Black:2023vju,Black:2024iwb} for first steps,
will be crucial to enhance the sensitivity of lifetime observables to NP effects and solidify their role in the search for BSM physics.

\section*{Acknowledgements}
This research was supported by the Deutsche Forschungsgemeinschaft (DFG, German Research Foundation) under grant 396021762 - TRR 257 and under grant 533766364 - EXC 3107/1. MLP acknowledges funding from the Dutch Research Council (NWO).
AR acknowledges the support by the Deutsche Forschungsgemeinschaft (DFG, German Research
Foundation) - project number 541305755.
The authors thank Daniel Moreno for helpful discussions concerning the results of Ref.\ \cite{Mannel:2020fts}.
ML\ would like to thank Joshua Davies, Manuel Egner, and Pascal Reeck for fruitful discussions.

\clearpage
\appendix
\section{\boldmath Computation in Dimensional Regularisation}
\label{sec:DR}
As a cross-check we have also performed the entire calculation in DR.
This avoids the need for introducing a light-quark mass as an infrared regulator, the role of which is taken as usual by $\epsilon = (4 - D)/2$.
We can thus set $m_u = m_d = m_s = 0$.
The downside is that the closed internal quark loop with $\gamma_5$ becomes troublesome.
One way to tackle this issue is the application of a Fierz transformation of the hermitian-conjugate four-fermion vertices, where one has to ensure validity of these transformations order by order in $\alpha_s$ by a suitable choice of evanescent operators, cf.\ e.g.\ Ref.~\cite{Egner:2024azu}.
Since we are working at leading order in $\alpha_s$ here, however, the definition of evanescent operators is relatively straightforward, see below.
One can then treat the remaining open spin line containing $\gamma_5$ in naive dimensional regularisation.
Using $\left\{\gamma_\mu, \gamma_5 \right\} = 0$ and $\gamma_5^2 = \mathbb{1}$ only Dirac chains with zero or one $\gamma_5$ remain.
The latter case does not contribute to the parity-even forward-scattering matrix element between external $B$-meson states and can therefore be dropped.
Applying the trace formalism of Refs.~\cite{Dassinger:2006md, Mannel:2023yqf} then closes the remaining chain of $\gamma$ matrices to a trace which, in absence of $\gamma_5$, is well-defined in DR.
The drawback is that, in contrast to the calculation in four dimensions, there is no Dirac trace over the $\bar{q}_2 q_3$ spin lines that can be evaluated early; instead one either has to keep longer Dirac chains until the last step or manually reduce them using identities such as $\gamma_\mu \gamma_\nu \gamma^\mu = (2 - D) \gamma_\nu$ and generalisations to more indices.\\
For the generation of Feynman diagrams we use \texttt{qgraf} \cite{Nogueira:1991ex}, the output of which is then processed into symbolic expressions and further manipulated using \texttt{tapir} \cite{Gerlach:2022qnc}, which we also use for the identification and minimisation of integral families.
Most of the remaining steps, including expansion of the propagators in the soft gluon background field,\footnote{The presence of problematic (in DR) $\gamma_5$ matrices and Levi-Civita tensors in \cref{eq:S1p} is spurious, since the quark propagator remains parity-conserving in presence of interactions with gluons. It is straightforward, at the expense of having more $\gamma$ matrices, to obtain an expression independent of these objects, see e.g.\ Ref.~\cite{Mannel:2020fts}.} Dirac algebra, tensor reduction, expansion around $p_b^2 \approx m_b^2$ in terms of covariant derivatives and identification of the effective operators defined in \cref{eq:hadronic_parameters} are then performed using custom \texttt{FORM} \cite{Vermaseren:2000nd,Kuipers:2012rf,Ruijl:2017dtg} routines, some of which are part of the ``Karlsruhe \texttt{calc}'' setup, used for example in Ref.~\cite{Davies:2025ghl}.
In the expansion of the internal light quarks in the soft-gluon background field we have not distinguished between specific operators and their colour structures.
Instead, for each combination of $\abs{\Delta B} = 1$ operators $\mathcal{Q}_i^{(q_1 q_2 q_3)\, (\prime)} \mathcal{Q}_j^{\dagger \, (q_1 q_2 q_3) (\prime)}$ we have expanded all three internal quark lines.
Subsequently, the evaluation of the colour trace and the power counting in $1/m_b$ ensure that only the appropriate emissions survive.
This approach of ``indifference'' comes at a slightly higher computational cost, but allows for a stronger cross-check of the more differentiated approach followed in four dimensions.
Finally, the list of scalar integrals is reduced to a small number of master integrals using \texttt{Kira} \cite{Maierhofer:2017gsa,Klappert:2020nbg}.
These $D$-dimensional master integrals can be expressed in terms of hypergeometric ${}_{2}{F}_1$ functions depending on $m_c^2/ p_b^2$; the relevant integrals for the most complicated case $b \to c \bar{c} s$ are presented in Ref.~\cite{Mannel:2020fts}.
The resulting expressions are then expanded in $\epsilon$ using \texttt{HypExp} \cite{Huber:2005yg,Huber:2007dx}.\\
In the process of renormalisation, the results for the $\Delta B = 0$ Wilson coefficients depend on the choice of evanescent operators.
We define the relevant operators as
\begin{align}
    \mathcal{E}_1 &= 4 (\bar{b} \gamma_{\mu\nu\rho} P_L q ) (\bar{q} \gamma^{\rho\nu\mu} P_L b) - (4 + a_1 \epsilon) \mathcal{O}_1 \,, \\
    \mathcal{E}_2 &= 4 (\bar{b} \gamma_{\mu\nu} P_L q ) (\bar{q} \gamma^{\nu\mu} P_L b) - (4 + a_2 \epsilon) \mathcal{O}_2 \,,
\end{align}
\begin{align}
    \mathcal{E}_3 &= 4 (\bar{b} \gamma_{\mu\nu\rho} P_L t^a q ) (\bar{q} \gamma^{\rho\nu\mu} P_L t^a b) - (4 + a_1 \epsilon) \mathcal{O}_3 \,, \\
    \mathcal{E}_4 &= 4 (\bar{b} \gamma_{\mu\nu} P_L t^a q ) (\bar{q} \gamma^{\nu\mu} P_L t^a b) - (4 + a_2 \epsilon) \mathcal{O}_4 \,, \\
    \mathcal{E}_5 &= 4 (\bar{b} \gamma_{\mu\nu\rho\sigma\tau} P_L q ) (\bar{q} \gamma^{\tau\sigma\rho\nu\mu} P_L b) - (16 + a_3 \epsilon) \mathcal{O}_1 \,, \\
    \mathcal{E}_6 &= 4 (\bar{b} \gamma_{\mu\nu\rho\sigma\tau} P_L t^a q ) (\bar{q} \gamma^{\tau\sigma\rho\nu\mu} P_L t^a b) - (16 + a_3 \epsilon) \mathcal{O}_3 \,, \\
    \mathcal{E}_7 &= 4 (\bar{b} \gamma_{\mu\nu\rho\sigma} P_L q ) (\bar{q} \gamma^{\sigma\rho\nu\mu} P_R b) - a_4 \epsilon \mathcal{O}_2 \,, \\
    \mathcal{E}_8 &= 4 (\bar{b} \gamma_{\mu\nu\rho\sigma} P_L t^a q ) (\bar{q} \gamma^{\sigma\rho\nu\mu} P_R t^a b) - a_4 \epsilon \mathcal{O}_4 \,,
\end{align}
where $\gamma_{\mu_1 \cdots \mu_n} \equiv \gamma_{\mu_1} \cdots \gamma_{\mu_n}$.
For the right-handed $\Delta B = 0$ operators $\mathcal{O}_{i}^{\prime}$ the corresponding evanescent operators are defined analogously.
Since only the operators $\mathcal{O}_{1,\ldots,4}^{(\prime)}$ contribute to the renormalisation of the Darwin term, no further evanescent operators are needed.
The renormalised Darwin term coefficients quoted in this work correspond to the choice $a_1 = a_2 = -8, a_3 = -64, a_4 = -24$.
This choice ensures that the renormalised coefficient functions obey $\mathcal{F}_{i,j}^{(q_1 q_2 q_3)} = \mathcal{F}_{j,i}^{(q_1 q_2 q_3)}$ for all $i$ and $j$.
Similarly to the computation of the bare two-loop two-quark amplitude, in the calculation of the matching of
\begin{equation}
    {\cal T} = i \int d^4 x \, T\, \{ {\cal H}_{\rm eff}(x), {\cal H}_{\rm eff}(0)\}
    \label{eq:tansop}
\end{equation}
to the effective $\Delta B = 0$ Hamiltonian relevant for the spectator-quark effects we have not distinguished between the different spectator topologies or decay channels.
Instead we have computed them fully automatically, summing over all effective $\abs{\Delta B} = 1$ operators and the quark flavours, for $B^+, B^0_d$, and $B^0_s$, separately.
The sum of these three contributions is then inserted into \cref{fig:MatrixMix}, where the terms proportional to each specific combination of CKM-matrix elements $\abs{V_{q_1 b}}^2 \abs{V_{q_2 q_3}}^2$ properly renormalise the corresponding bare partial $b \to q_1 \bar{q}_2 q_3$ two-loop amplitude.
Finally, the renormalised one-loop matrix elements of the four-fermion $\Delta B = 0$ operators in the soft background gluon field, see \cref{fig:MatrixMix}, are given by a relation similar to \cref{eq:crD-finite}, with the replacements $\log (m_q^2 / m_b^2) \to 1/\epsilon$ and $b_{1,2} \to \tilde{b}_{1,2}$, where $\tilde{b}_{1,2} = -2/3$.

\section{\texorpdfstring{\boldmath Results for the $b \to u \bar c s$ transition}{Results for the b -> u cbar s transition}}\label{sec:resultsucs}
\label{sec:ucs}
In this appendix, we present the results for the CKM-suppressed decay channel $b \to u \bar{c} s$.
We emphasize again that these contributions require the calculation not only of the WE and PI topologies, performed in Ref.~\cite{Lenz:2022pgw} but also of the WA diagrams.
Results for the latter are listed in \cref{sec:WA}.
\paragraph{Dimension-three contribution:}
The independent functions $P_{i,j}^{ucs}(\rho)$ in \cref{eq:Fij} are given by
\begin{align}
P_{1,1}^{(ucs)}(\rho) &= 
1 - 8\rho - 12\rho^2 \log\rho + 8\rho^3 - \rho^4 \,, \label{eq:P11ucs} \\[2mm]
P_{1,7}^{(ucs)}(\rho) &= \sqrt{\rho} \big(-1 - 9\rho + 9\rho^2 + \rho^3 - 6\rho(1 + \rho) \log\rho \big) \,. \label{eq:P17ucs}
\end{align}
The other coefficients follow from the relations
\begin{equation}
\begin{aligned}
& P_{2,2}^{(ucs)} = P_{3,3}^{(ucs)} = P_{4,4}^{(ucs)} = 4P_{5,5}^{(ucs)} = 4P_{6,6}^{(ucs)} = 4P_{7,7}^{(ucs)} = 4P_{8,8}^{(ucs)} = \frac{1}{12}P_{9,9}^{(ucs)} = \frac{1}{12}P_{10,10}^{(ucs)} = P_{1,1}^{(ucs)} \,, \\[6pt]
& P_{3,4}^{(ucs)} = 4P_{5,6}^{(ucs)} = 4P_{7,8}^{(ucs)} = \frac{1}{12}P_{9,10}^{(ucs)} = P_{1,1}^{(ucs)} \,, \\[6pt]
& P_{1,8}^{(ucs)} = P_{2,8}^{(ucs)} = -\frac{1}{12}P_{1,9}^{(ucs)} = -\frac{1}{12}P_{1,10}^{(ucs)} = -\frac{1}{12}P_{2,10}^{(ucs)} = P_{3,5}^{(ucs)} = P_{4,5}^{(ucs)} = P_{1,7}^{(ucs)} \,, \\[6pt]
& P_{i,j}^{(ucs)} = P_{i-1,j+1}^{(ucs)} \quad \text{for even } i \text{ and odd } j \,,
\end{aligned}
\label{eq:relations_ucsdim3}
\end{equation}
with all remaining combinations vanishing.

\paragraph{Dimension-five contribution:}
For the contribution of the chromomagnetic operator, the independent non-vanishing functions $G_{i,j}^{ucs}(\rho)$ in \cref{eq:Fij} read:
\begin{align}
G_{1,1}^{(ucs)}(\rho) &= 
-\frac{1}{2}\big(5\rho^4 - 24\rho^3 + 24\rho^2 + 12\rho^2\log\rho - 8\rho + 3\big) \,, \label{eq:G11ucs} \\[2mm]
G_{1,2}^{(ucs)}(\rho) &= -\frac{1}{2} \big(19 + 16\rho - 24\rho^2 - 16\rho^3 + 5\rho^4 + 12\rho(4 + \rho)\log\rho\big) \,, \label{eq:G12ucs} \\[2mm]
G_{1,7}^{(ucs)}(\rho) &= \frac{1}{2} \sqrt{\rho} \big(-5 + 3\rho - 3\rho^2 + 5\rho^3 - 6\rho(1+\rho)\log\rho\big) \,, \label{eq:G17ucs} \\[2mm]
G_{1,8}^{(ucs)}(\rho) &= \frac{3}{2} \sqrt{\rho} \big(-9 + 7\rho + \rho^2 + \rho^3 - 2(2 + 3\rho + \rho^2)\log\rho\big) \,, \label{eq:G18ucs} \\[2mm]
G_{1,9}^{(ucs)}(\rho) &= 2 \sqrt{\rho} \big(-17 + 15\rho + 9\rho^2 - 7\rho^3 + 6\rho(-5 + 3\rho)\log\rho\big) \,, \label{eq:G19ucs} \\[2mm]
G_{2,8}^{(ucs)}(\rho) &= \frac{1}{2} \sqrt{\rho} \big(11 - 9\rho - 3\rho^2 + \rho^3 - 6(-3 + \rho)\rho\log\rho\big) \,, \label{eq:G28ucs} \\[2mm]
G_{2,10}^{(ucs)}(\rho) &= -2 \sqrt{\rho} \big(1 - 3\rho - 9\rho^2 + 11\rho^3 + 6(1 - 3\rho)\rho\log\rho\big) \,, \label{eq:G210ucs} \\[2mm]
G_{3,4}^{(ucs)}(\rho) &= \frac{1}{2} \big(13 - 40\rho + 24\rho^2 + 8\rho^3 - 5\rho^4 - 12\rho^2\log\rho\big) \,, \label{eq:G34ucs} \\[2mm]
G_{3,6}^{(ucs)}(\rho) &= \frac{1}{6} \sqrt{\rho} \big(-17 - 9\rho + 9\rho^2 + 17\rho^3 - 6(-2 + 9\rho + 3\rho^2)\log\rho\big) \,, \label{eq:G36ucs} \\[2mm]
G_{7,10}^{(ucs)}(\rho) &= -4 \big(-4 + 3\rho + \rho^3 - 6\rho\log\rho\big) \,, \label{eq:G710ucs} \\[2mm]
G_{8,10}^{(ucs)}(\rho) &= 8 (1 - \rho)^3 \,, \label{eq:G810ucs} \\[2mm]
G_{9,9}^{(ucs)}(\rho) &= -2 \big(25 - 72\rho + 120\rho^2 - 88\rho^3 + 15\rho^4 + 36\rho^2\log\rho\big) \,, \label{eq:G99ucs} \\[2mm]
G_{9,10}^{(ucs)}(\rho) &= -2 \big(25 + 72\rho - 72\rho^2 - 40\rho^3 + 15\rho^4 + 12\rho(8 + 3\rho)\log\rho\big) \,, \label{eq:G910ucs} \\[2mm]
G_{10,10}^{(ucs)}(\rho) &= 2 \big(23 - 72\rho + 24\rho^2 + 40\rho^3 - 15\rho^4 - 36\rho^2\log\rho\big) \,. \label{eq:G1010ucs}
\end{align}
The remaining non-vanishing coefficient functions follow from the relations
\begin{equation}
\begin{aligned}
& G_{2,2}^{(ucs)} = G_{3,3}^{(ucs)} = 4G_{6,6}^{(ucs)} = 4G_{8,8}^{(ucs)} = G_{1,1}^{(ucs)} \,, \qquad G_{1,10}^{(ucs)} = -12G_{1,8}^{(ucs)} \,, \\[6pt]
& G_{4,4}^{(ucs)} = 4G_{5,5}^{(ucs)} = 4G_{5,6}^{(ucs)} = 4G_{7,7}^{(ucs)} = 4G_{7,8}^{(ucs)} = G_{3,4}^{(ucs)} \,, \qquad G_{3,5}^{(ucs)} = G_{4,6}^{(ucs)} = G_{1,7}^{(ucs)} \,, \\[6pt]
& G_{i,j}^{(ucs)} = G_{i-1,j+1}^{(ucs)} \quad \text{for even } i \text{ and odd } j \,,
\end{aligned}
\label{eq:relations_ucsdim5}
\end{equation}
with all remaining combinations vanishing.
\paragraph{Dimension-six contribution:}
For the contribution of the Darwin operator, the independent non-vanishing functions $D_{i,j}^{ucs}(\rho)$ in \cref{eq:Fij} read:
\begin{align}
D_{1,1}^{(ucs)}(\rho) &= \frac{2}{3}(1-\rho) \big(9 + 11\rho - 25\rho^2 + 5\rho^3 - 24(1 - \rho^2)\log(1-\rho) - 12\rho^2\log\rho\big) \,, \label{eq:D11ucs} \\[2mm]
D_{1,2}^{(ucs)}(\rho) &= -\frac{2}{3} \big(41 - 26\rho + 18\rho^2 - 38\rho^3 + 5\rho^4 + 48(1-\rho)^2(1 + \rho)\log(1-\rho) \nonumber \\
&\quad + 12(2 + 5\rho + 2\rho^2 - 2\rho^3)\log\rho\big) \,, \label{eq:D12ucs} \\[2mm]
D_{1,7}^{(ucs)}(\rho) &= \frac{2}{3} \sqrt{\rho} \big(-8 + 27\rho - 24\rho^2 + 5\rho^3 + 12(1-\rho)^2\log(1-\rho) - 6\rho(-2 + \rho)\log\rho\big) \,, \label{eq:D17ucs} \\[2mm]
D_{1,8}^{(ucs)}(\rho) &= \frac{2}{3} \sqrt{\rho} \big(7 + 6\rho - 15\rho^2 + 2\rho^3 + 24(1 - \rho)^2\log(1-\rho) + 6(1 + 4\rho - 2\rho^2)\log\rho\big) \,, \label{eq:D18ucs} \\[2mm]
D_{1,9}^{(ucs)}(\rho) &= -8 \sqrt{\rho} \big(\rho(7 - 8\rho + \rho^2 - 6(-2 + \rho)\log\rho) + 12(1-\rho)^2\log(1-\rho)\big) \,, \label{eq:D19ucs} \\[2mm]
D_{2,8}^{(ucs)}(\rho) &= -\frac{2}{3} \sqrt{\rho} \big(-4 + 3\rho + \rho^3 - 12(1 - \rho)^2\log(1-\rho) + 6\rho(-2 + \rho)\log\rho\big) \,, \label{eq:D28ucs} \\[2mm]
D_{2,10}^{(ucs)}(\rho) &= -8 \sqrt{\rho} \big(-4 + 17\rho - 16\rho^2 + 3\rho^3 + 12(1-\rho)^2\log(1-\rho) - 6\rho(-2+\rho)\log\rho\big) \,, \label{eq:D210ucs} \\[2mm]
D_{3,4}^{(ucs)}(\rho) &= -\frac{2}{3} \big(-51 + 94\rho - 42\rho^2 - 6\rho^3 + 5\rho^4 + 24(1 - \rho)^2(3 + \rho)\log(1-\rho) \nonumber \\
&\quad - 12(-1 + \rho^2 + \rho^3)\log\rho\big) \,, \label{eq:D34ucs} \\[2mm]
D_{3,6}^{(ucs)}(\rho) &= 4 \sqrt{\rho} \big(1 + 3\rho - 5\rho^2 + \rho^3 + 4(1-\rho)^2\log(1-\rho) + 2(1 + 2\rho - \rho^2)\log\rho\big) \,, \label{eq:D36ucs} \\[2mm]
D_{4,4}^{(ucs)}(\rho) &= -\frac{2}{3} \big(-59 + 112\rho - 66\rho^2 + 8\rho^3 + 5\rho^4 + 48(1-\rho)^2\log(1-\rho) - 24\rho^2\log\rho\big) \,, \label{eq:D44ucs} \\[2mm]
D_{7,8}^{(ucs)}(\rho) &= -\frac{1}{6} \big(-55 + 106\rho - 54\rho^2 - 2\rho^3 + 5\rho^4 + 24(1 - \rho)^2(3+\rho)\log(1-\rho) \nonumber \\
&\quad - 12(-1 + \rho^2 + \rho^3)\log\rho\big) \,, \label{eq:D78ucs} \\[2mm]
D_{7,10}^{(ucs)}(\rho) &= \frac{8}{3} \big(22 - 27\rho + 12\rho^2 - 7\rho^3 - 6(1-\rho)^3\log(1-\rho) + 3(1 + 5\rho + 3\rho^2 - \rho^3)\log\rho\big) \,, \label{eq:D710ucs} \\[2mm]
D_{8,8}^{(ucs)}(\rho) &= \frac{1}{6} (1-\rho) \big(13 + 3\rho - 21\rho^2 + 5\rho^3 - 24(1-\rho^2)\log(1-\rho) - 12\rho^2\log\rho\big) \,, \label{eq:D88ucs} \\[2mm]
D_{8,10}^{(ucs)}(\rho) &= \frac{4}{3} \big(23 - 51\rho + 45\rho^2 - 17\rho^3 - 12(1-\rho)^3\log(1-\rho) + 6(3 - \rho)\rho^2\log\rho\big) \,, \label{eq:D810ucs} 
\end{align}
\begin{align}
D_{9,9}^{(ucs)}(\rho) &= -\frac{8}{3} \big(7 - 72\rho + 162\rho^2 - 112\rho^3 + 15\rho^4 + 48(1-\rho)^2(1+2\rho)\log(1-\rho) \nonumber \\
&\quad + 24(3 - 2\rho)\rho^2\log\rho\big) \,, \label{eq:D99ucs} \\[2mm]
D_{9,10}^{(ucs)}(\rho) &= -\frac{8}{3} \big(11 + 102\rho - 66\rho^2 - 62\rho^3 + 15\rho^4 + 24(1-\rho)^2(7+5\rho)\log(1-\rho) \nonumber \\
&\quad + 12(5 + 10\rho + 3\rho^2 - 5\rho^3)\log\rho\big) \,, \label{eq:D910ucs} \\[2mm]
D_{10,10}^{(ucs)}(\rho) &= -\frac{8}{3} \big(-131 + 234\rho - 108\rho^2 - 10\rho^3 + 15\rho^4 + 24(1-\rho)^2(5+\rho)\log(1-\rho) \nonumber \\
&\quad - 12\rho^2(3 + \rho)\log\rho\big) \,. \label{eq:D1010ucs}
\end{align}
The remaining non-vanishing coefficient functions follow from the relations
\begin{equation}
\begin{aligned}
& D_{2,2}^{(ucs)} = D_{3,3}^{(ucs)} = 4D_{6,6}^{(ucs)} = D_{1,1}^{(ucs)} \,, \qquad D_{1,10}^{(ucs)} = -12D_{1,8}^{(ucs)} \,, \qquad D_{3,5}^{(ucs)} = D_{4,6}^{(ucs)} = D_{1,7}^{(ucs)} \,, \\[6pt]
& D_{5,5}^{(ucs)} = D_{7,7}^{(ucs)} = \frac{1}{4}D_{4,4}^{(ucs)} \,, \qquad D_{5,6}^{(ucs)} = \frac{1}{4}D_{3,4}^{(ucs)} \,, \\[6pt]
& D_{i,j}^{(ucs)} = D_{i-1,j+1}^{(ucs)} \qquad \text{for even } i \text{ and odd } j \,.
\end{aligned}
\label{eq:relations_ucsdim6}
\end{equation}
\section{\texorpdfstring{\boldmath Results for the $b \to u \bar u d$ transition}{Results for the b -> u ubar d transition}}\label{sec:resultsuud}
\label{sec:uud}
The dimension-three and five contributions to the $b \to u\bar u d$ channel are obtained directly from the corresponding expressions for the $b \to c \bar u d$ transition by taking the massless limit $\rho \to 0$.
In contrast, the coefficient of the Darwin operator requires a dedicated calculation to properly regularise the IR divergences originating in this case from the expansion of the $q_1$-quark propagator. The corresponding results are presented below.

\paragraph{Dimension-six contribution:}
The independent non-vanishing functions $D_{i,j}^{(uud)}(\rho)$ for the coefficient of the Darwin operator in \cref{eq:Fij} read:
\begin{align}
D_{1,1}^{(uud)}(\rho) &= 6 \,, & D_{1,2}^{(uud)}(\rho) &= -\frac{34}{3} \,, & D_{3,4}^{(uud)}(\rho) &= \frac{110}{3} \,, & D_{4,4}^{(uud)}(\rho) &= \frac{118}{3} \,, \label{eq:D11uud} \\[2pt]
D_{5,5}^{(uud)}(\rho) &= \frac{59}{6} \,, & D_{5,6}^{(uud)}(\rho) &= \frac{55}{6} \,, & D_{6,6}^{(uud)}(\rho) &= \frac{3}{2} \,, & D_{7,8}^{(uud)}(\rho) &= \frac{21}{2} \,, \label{eq:D55uud} \\[2pt]
D_{7,10}^{(uud)}(\rho) &= 48 \,, & D_{8,8}^{(uud)}(\rho) &= \frac{13}{6} \,, & D_{8,10}^{(uud)}(\rho) &= \frac{92}{3} \,, & D_{9,9}^{(uud)}(\rho) &= -\frac{56}{3} \,, \label{eq:D710uud} \\[2pt]
D_{9,10}^{(uud)}(\rho) &= 120 \,, & D_{10,10}^{(uud)}(\rho) &= \frac{1048}{3} \,. & & & & \label{eq:D910uud}
\end{align}
The remaining non-vanishing coefficients follow from the relations
\begin{align}
D_{3,3}^{(uud)} &= D_{1,1}^{(uud)} \,, \qquad D_{7,7}^{(uud)} = D_{5,5}^{(uud)} \,, \qquad D_{2,2}^{(uud)} = D_{1,1}^{(uud)} \,, \nonumber
\\[6pt]
D_{i,j}^{(uud)} & = D_{i-1,j+1}^{(uud)} \qquad \quad \text{for even } i \text{ and odd } j. 
\label{eq:relations_uud}
\end{align}

\section{\boldmath Results for the WA Diagrams} 
\label{sec:WA}
In the presence of the generalised operator basis in the effective Hamiltonian in
\cref{eq:Heff-NP}, the transition operator defined in \cref{eq:tansop}, which enters \cref{eq:GammaB}, 
receives additional contributions from four-quark operator topologies, namely PI, WE, and WA, see \cref{fig:4QDiags}.  
The PI and WE contributions generated by the BSM Hamiltonian were already computed in 
Ref.~\cite{Lenz:2022pgw}.  
To complete the set of four-quark operator contributions, we derive here the corresponding expressions for the WA topology.\\
Following the notation of Ref.~\cite{Lenz:2022pgw}, the BSM contribution to the transition operator
for the WA topology and internal quark flavours $q_2 = c, u$, and $q_3 = d, s$ can be written as
\begin{equation}
\mathrm{Im}\,\mathcal{T}_{\mathrm{WA}}^{q_2 q_3}
=
\frac{G_F^2 m_b^2}{6\pi}\,|V_{ub}|^2 |V_{q_{2}q_{3}}|^2\, (1-\rho)^2\,
\bigg[
\sum_{m,n = 1}^{20}
{\cal{C}}_m\, {\cal{C}}_n^{*}\, {\mathcal{N}}^{\prime}_{mn}\, A^{\mathrm{WA},\, q_2 q_3}_{mn}
\bigg] \,,
\label{eq:T-NP}
\end{equation}
where $\mathcal{N}^{\prime}_{mn}$ denotes the colour factor, which takes the value $N_c$ if both $m$ and $n$ are odd, and 1 otherwise, and 
$A^{\mathrm{WA},\, q_2 q_3}_{mn}$ are the coefficient functions.
For brevity, primed Wilson coefficients in \cref{eq:Heff-NP} are absorbed into the same index via
\begin{equation}
{\cal{C}}_m \equiv {\cal{C}}^{\prime\,}_{\,m-10}\,, \qquad m>10 \,.
\end{equation}

In the following we present explicit results for the coefficient functions $A^{{\rm WA},\, q_2 q_3}_{mn}$ in \cref{eq:T-NP}. 
We provide results for the case $q_2 q_3 = c s$, retaining the full charm-mass dependence with $\rho = m_c^2/m_b^2$. 
The corresponding results for the massless case $q_2 q_3 = ud$ are obtained by taking the limit $\rho \to 0$.
\begin{align}
A^{\rm WA, c s}_{1,1} &= -\frac{1}{2} \,\bigl[ \mathcal{O}_1 (2 + \rho) - 2\, \mathcal{O}_2 (1 + 2 \rho) \bigr] \, , \quad 
A^{\rm WA, c s}_{1,3} = -\frac{1}{2} \,  \bigl[ \mathcal{O}_5' (2 + \rho) - 2 \,\mathcal{O}_6' (1 + 2 \rho) \bigr]\, , \\
A^{\rm WA, c s}_{1,5} &= -\frac{3}{2} \, \mathcal{O}_6' \,\sqrt{\rho}\, , \quad 
A^{\rm WA, c s}_{1,7} = -\frac{3}{2}  \,\mathcal{O}_2 \,\sqrt{\rho}\, , \quad 
A^{\rm WA, c s}_{1,9} = -6 \sqrt{\rho} \bigl[\mathcal{O}_1 - \mathcal{O}_2 \bigr] \,, \\
A^{\rm WA, c s}_{3,5} &= -\frac{3}{2}\,  \mathcal{O}_2' \, \sqrt{\rho} \, ,\quad 
A^{\rm WA, c s}_{3,7} = -\frac{3}{2}  \, \mathcal{O}_6 \, \sqrt{\rho}\, , \quad 
A^{\rm WA, c s}_{3,9} = -6 \sqrt{\rho} \bigl[ \mathcal{O}_5 - \mathcal{O}_6 \bigr]\,,\\ 
A^{\rm WA, c s}_{5,5} &= \frac{3}{2}\, \mathcal{O}_2' \, ,\quad 
A^{\rm WA, c s}_{5,7} = \frac{3}{2}\, \mathcal{O}_6 \, ,\quad 
A^{\rm WA, c s}_{9,9} = -8 \, (\mathcal{O}_1 - \mathcal{O}_2) \, (1 + 2 \rho) \,,
\end{align}
where the $\Delta B = 0$ operators ${\cal O}_i^{(\prime)}$ are defined in \cref{eq:O1-O2-HQET,eq:O3-O4-HQET,eq:O5-O6-HQET,eq:O7-O8-HQET}.
The other non-vanishing coefficients can be obtained from the symmetry relations
\begin{align}
&A^{{\rm WA}, cs}_{3,3} = A^{{\rm WA}, cs}_{1,1}\Big|_{\mathcal O_i \leftrightarrow \mathcal O'_i}\,, \qquad 
A^{{\rm WA}, cs}_{7,7} = A^{{\rm WA}, cs}_{5,5}\Big|_{\mathcal O_i \leftrightarrow \mathcal O'_i}\,, \\[6pt]
&A^{{\rm WA}, cs}_{m,n} = 
\begin{cases}
A^{{\rm WA}, cs}_{m,n-1} & (m \text{ odd}, n \text{ even}) \\[4pt]
A^{{\rm WA}, cs}_{m-1,n+1} & (m \text{ even}, n \text{ odd}) 
\end{cases} \,, \\[6pt]
& A^{{\rm WA},cs}_{m, \, n} = A^{{\rm WA},cs}_{m - 1, \,n-1} 
\Big|_{{\cal{O}}^{(\prime)}_i \to \left(\frac{{\cal{O}}^{(\prime)}_i}{N_c} + 2 \, {\cal{O}}^{(\prime)}_{i+2}\right)}\, ,  \quad  (m,n \text{ even}) \,, \\
&A^{{\rm WA}, cs}_{m,n} = A^{{\rm WA}, cs}_{n,m}\Big|_{\mathcal O_i \leftrightarrow \mathcal O'_i} \,, \quad (m \neq n) \,,\\[6pt]
&A^{{\rm WA}, cs}_{m,n} = A^{{\rm WA}, cs}_{m+10,n+10}\Big|_{\mathcal O_i \leftrightarrow \mathcal O'_i}\,, \quad m,n \in \{1,\dots,10\}\,,
\end{align}
with all remaining combinations vanishing. 
\section{\boldmath Contribution of the QCD-penguin operators in the SM up to dimension-six
} 
\label{sec:SM_penguin}
Using the relations in \cref{eq:Q1_Q1cal,eq:O3_O4,eq:O5_O6}, the matching coefficients to the two-quark operator contributions to the HQE of a $B$ meson, up to dimension-six, originating from the insertion of the QCD-penguin operators $Q_3^{qq_3}, \ldots, Q_6^{qq_3}$ in \cref{eq:Heff-SM}, within the SM, can be straightforwardly derived from the expressions for the coefficients $F_{i,j}^{q_1 q_2 q_3}$ computed in this work. \\
The interference of the current-current operators $Q_{1,2}^{q q q_3}$ and of the penguin operators $Q_{3, \ldots, 6}^{qq_3}$ yields:
\begin{align}
\frac{\Gamma_{\rm PO}^{\rm int}(b \to q\bar qq_3)}{\Gamma_0} = 2  
\mbox{Re}\left(\xi_{q q_3} \xi_{tq_3}^* \right)\Biggl[ \left(C_1 C_3 \,\mathcal{F}_{1,2}^{(q q q_3)} + C_1 C_4 \,\mathcal{F}_{1,1}^{(q q q_3)}  + C_2 C_3 \,\mathcal{F}_{2,2}^{(q q q_3)}+ C_2 C_4 \,\mathcal{F}_{2,1}^{(q q q_3)}
\right)
\nonumber \\
-2 \left( C_1 C_5 \mathcal{F}_{1,6}^{(q q q_3)} + C_1 C_6 \mathcal{F}_{1,5}^{(q  q q_3)} + C_2 C_5 \mathcal{F}_{2,6}^{(q q q_3)}  + C_2 C_6 \mathcal{F}_{2,5}^{(q q q_3)}
\right)
\Biggr] \,,
\label{eq:Peng_mixed}
\end{align}
where it is $q = u,c,$ and $q_3 = d,s$, and we have defined $\xi_{x y} \equiv V_{x b} V_{x y}^*$.
The contribution arising from the quadratic insertion of the QCD-penguin operators $Q_{3, \ldots, 6}^{qq_3}$ in \cref{eq:GammaB} reads:
\begin{align}
\frac{\Gamma_{\rm PO}^{\rm quad}(b \to q \bar q q_3)}{\Gamma_0} = 
|\xi_{t q_3}|^2  \Bigg[   C_3^2 {\cal F}_{2,2}^{qqq_3} + 2 C_3 C_4 {\cal F}_{2,1}^{qqq_3} +  C_4^2 {\cal F}_{1,1}^{qqq_3} + 4 \left( C_5^2 {\cal F}_{6,6}^{qqq_3} + 2 C_5 C_6 {\cal F}_{6,5}^{qqq_3} + C^2_6 {\cal F}_{5,5}^{qqq_3}  \right)
\nonumber \\
- 4 \left( C_3 C_5 {\cal F}_{2,6}^{qqq_3} + C_3 C_6 {\cal F}_{2,5}^{qqq_3} + C_4 C_5 {\cal F}_{1,6}^{qqq_3} + C_4 C_6 {\cal F}_{1,5}^{qqq_3} \right)
\Bigg]\,.
\label{eq:Peng_quadratic}
\end{align}
Note that \cref{eq:Peng_quadratic} actually holds for $q = u,d,s,c$. In fact, for massless $u,d,s,$ quarks, we have: 
\begin{equation}
    {\cal F}_{i,j}^{ddq_3} =  {\cal F}_{i,j}^{ssq_3} =  {\cal F}_{i,j}^{uuq_3}\,.
\end{equation}
We have explicitly verified that, at dimension-three, inserting the expressions for the leading-power functions $P_{i,j}^{c c s}(\rho)$ in \cref{eq:Peng_mixed,eq:Peng_quadratic} reproduces the known results from Ref.~\cite{Krinner:2013cja} for the case of $b \to c \bar c s$ transition. 
Moreover, we emphasise that, restricting ourselves to the SM case, the Wilson coefficients appearing in the above equations are real-valued. \\
Substituting the explicit expressions for the functions ${\cal F}_{i,j}^{qqq_3}$ yields analytic results for the matching coefficients to the two-quark operator contributions to $\Gamma(B)$, in the HQE and up to dimension-six, due to the QCD-penguin operators. These results complete the contribution of the QCD-penguin operators to the total $B$-meson width within the HQE at this order.
The matching coefficients to the four-quark operator contributions were, in fact, already computed in Ref.~\cite{Franco:2002fc}. 

\bibliographystyle{JHEP}
\bibliography{citation}

\end{document}